\documentclass[longauth]{aa}  
\usepackage[utf8]{inputenc}
\usepackage{CJKutf8}

\usepackage{CJKutf8}

\usepackage{graphicx}
\usepackage{txfonts}
\usepackage{lipsum}
\usepackage{subcaption}         
\usepackage{lscape}             
\usepackage{placeins}           
\usepackage{xcolor}   
\usepackage{hyperref}   

\newcommand\omc{$\omega$\,Cen}

\addto\extrasenglish{
    
}

\addto\extrasenglish{
    
}

\begin{document}
   \title{SDSS-V Local Volume Mapper (LVM): The Integrated Light and Internal Rotation of Omega Centauri}
    \titlerunning{Omega Centauri studied with the SDSS-V Local Volume Mapper}
    \authorrunning{M. Häberle et al.} 

   \author{Maximilian Häberle\inst{\ref{inst_eso}}\corrauth{maximilian.haberle@eso.org}
    \and Dmitry Bizyaev\inst{\ref{inst_apo}}\email{dmbiz@apo.nmsu.edu}    
    \and Alina Boecker\inst{\ref{inst_uwien}}\email{alina.boecker@univie.ac.at}
    \and Callie Clontz\inst{\ref{inst_mpia}}\email{clontz@mpia.de}
    \and Bruno Dias\inst{\ref{inst_andresbello}}\email{astro.bdias@gmail.com}
    \and Antoine Dumont\inst{\ref{inst_mpia}}
    \and Evgeniya Egorova\inst{\ref{inst_ari}}\email{e.egorova@uni-heidelberg.de}
    \and Anja Feldmeier-Krause\inst{\ref{inst_uwien}}\email{anja.krause@univie.ac.at}
    \and Jos\'e G. Fern\'andez-Trincado\inst{\ref{inst_ubh}}\email{jose.fernandez@ucn.cl}
    \and Pablo García\inst{\ref{inst_ucn},\ref{naoc}}\email{pablo.garcia@ucn.cl}
    \and Thomas M. Herbst\inst{\ref{inst_mpia}}\email{herbst@mpia.de}
    \and Thomas Hilder\inst{\ref{inst_monash}}\email{thomas.hilder@monash.edu}
    \and Hector Javier Ibarra-Medel\inst{\ref{inst_unam}}\email{hjibarram@gmail.com}
    \and Amy  M. Jones\inst{\ref{inst_stsci}}\email{amjones@stsci.edu}
    \and Ralf Klessen\inst{\ref{inst_ita}}\email{klessen@uni-heidelberg.de}
    \and Nicholas P. Konidaris II\inst{\ref{inst_carnegie}}\email{npk@carnegiescience.edu}
    \and Kathryn Kreckel\inst{\ref{inst_ari}}\email{kathryn.kreckel@uni-heidelberg.de}
    \and Alejandra Z. Lugo-Aranda\inst{\ref{inst_unam_ensenada},\ref{inst_andresbello_concepcion}}\email{alugo@astro.unam.mx}
    \and Alfredo Mejía-Narváez\inst{\ref{inst_uc}}\email{alfredo@das.uchile.cl}
    \and Nadine Neumayer\inst{\ref{inst_mpia}}\email{neumayer@mpia.de}
    \and Hans-Walter Rix\inst{\ref{inst_mpia}}\email{rix@mpia.de}
    \and Alexandre Roman-Lopes\inst{\ref{inst_uni_la_serena}}\email{aroman@userena.cl}
    \and Sebastián Sánchez\inst{\ref{inst_unam}}\email{sfsanchez@astro.unam.mx}
    \and Saroon Sasi\inst{\ref{inst_andresbello}}\email{s.sasi@uandresbello.edu}
    \and Anil Seth\inst{\ref{inst_utah}}\email{aseth@astro.utah.edu}
    \and Amrita Singh\inst{\ref{inst_uc}}\email{amrita@das.uchile.cl}
    \and Peter Smith\inst{\ref{inst_mpia},\ref{inst_hgsfp}}\email{pesmith@mpia.de}
    \and Rodolfo de J. Zerme\~{n}o\inst{\ref{inst_unam}}\email{rzermeno@astro.unam.mx}
    \and Zixian Wang (\begin{CJK}{UTF8}{gbsn}王梓先)\end{CJK}\inst{\ref{inst_utah}}\email{zixian.wang@utah.edu}
    \and Guillermo A. Blanc\inst{\ref{inst_carnegie},\ref{inst_uc}}\email{gblancm@carnegiescience.edu}
    \and Joel R. Brownstein\inst{\ref{inst_utah}}\email{joelbrownstein@astro.utah.edu}
    \and Niv Drory\inst{\ref{inst_mcdonald}}\email{drory@astro.as.utexas.edu}
    \and Oleg Egorov\inst{\ref{inst_ari}}\email{oleg.egorov@uni-heidelberg.de}
    \and Evelyn J. Johnston\inst{\ref{inst_diegoportales}}\email{evelynjohnston.astro@gmail.com}
    \and Ivan Katkov\inst{\ref{inst_nyuad},\ref{inst_sternberg}}\email{ik52@nyu.edu}
    \and Juna A. Kollmeier\inst{\ref{inst_carnegie}}\email{jak@carnegiescience.edu}
    \and Sorya Lambert\inst{\ref{inst_diegoportales}}\email{sorya.lambert@mail.udp.cl}
}
    \institute{European Southern Observatory, Karl-Schwarzschild-Straße 2, 85748 Garching, Germany\label{inst_eso}
    \and Apache Point Observatory, P.O. Box 59, Sunspot, NM 88349, USA\label{inst_apo}
    \and The Observatories of the Carnegie Institution for Science, 813 Santa Barbara Street, Pasadena, CA 91101, USA\label{inst_carnegie}
    \and Departamento de Astronom\'{i}a, Universidad de Chile, Camino del Observatorio 1515, Las Condes, Santiago, Chile\label{inst_uc}
    \and Department of Astrophysics, University of Vienna, T\"urkenschanzstrasse 17, 1180 Wien, Austria\label{inst_uwien}
    \and Max Planck Institute for Astronomy, K\"onigstuhl 17, D-69117 Heidelberg, Germany\label{inst_mpia}
    \and Instituto de Astrof\'isica, Departamento de F\'isica y Astronom\'ia, Facultad de Ciencias Exactas, Universidad Andres Bello, Fernandez Concha, 700, Las Condes, Santiago, Chile\label{inst_andresbello}
    \and Institute of Astrophysics, Facultad de Ciencias Exactas, Universidad Andrés Bello, Sede Concepción, Talcahuano, Chile\label{inst_andresbello_concepcion}
    \and McDonald Observatory, The University of Texas at Austin, 1 University Station, Austin, TX 78712, USA\label{inst_mcdonald}
    \and Astronomisches Rechen-Institut, Zentrum f\"{u}r Astronomie der Universit\"{a}t Heidelberg, M\"{o}nchhofstr. 12-14, D-69120 Heidelberg, Germany\label{inst_ari}
    \and Centro de investigación en Astronomía, Facultad de Ingeniería, Ciencia y Tecnología, Universidad Bernardo O’Higgins, Av. Viel 1497, Santiago, 8370993, Chile\label{inst_ubh}
    \and Instituto de Astronom\'{i}a, Universidad Cat\'{o}lica del Norte, Av. Angamos 0610, Antofagasta, Chile\label{inst_ucn}
    \and Chinese Academy of Sciences South America Center for Astronomy, National Astronomical Observatories, CAS, Beijing 100101, China\label{naoc}
    \and School of Physics \& Astronomy, Monash University, Wellington Road, Clayton, Victoria 3800, Australia\label{inst_monash}
    \and Instituto de Astronom\'{i}a, Universidad Nacional Aut\'{o}noma de M\'{e}xico, A.P. 70-264, 04510, Mexico, D.F., M\'{e}xico\label{inst_unam}
    \and Universidad Nacional Autónoma de México. Instituto de Astronomía. A.P. 106, 22800. Ensenada, B.C., México\label{inst_unam_ensenada}
    \and Universidad Diego Portales, Instituto de Estudios Astrof\'{i}sicos, Facultad de Ingenier\'{i}a y Ciencias, Av. Ej\'{e}rcito Libertador 441, Santiago, Chile\label{inst_diegoportales}
    \and New York University Abu Dhabi, PO Box 129188, Abu Dhabi, UAE\label{inst_nyuad}
    \and Sternberg Astronomical Institute, Lomonosov Moscow State University, Universitetskij pr. 13, 119234 Moscow, Russia\label{inst_sternberg}
    \and Institut f\"{u}r theoretische Astrophysik, Zentrum f\"{u}r Astronomie der Universit\"{a}t Heidelberg, Albert-Ueberle-Str. 2, D-69120 Heidelberg, Germany\label{inst_ita}
    \and Department of Physics and Astronomy, University of Utah, 270 S. 1400 E. \#E2108, Salt Lake City, UT 84112, USA\label{inst_utah}
    \and Department of Astronomy, Universidad de La Serena, Av. Raul Bitran 1302, La Serena, Chile\label{inst_uni_la_serena}
    \and Space Telescope Science Institute, 3700 San Martin Drive, Baltimore, MD 21218, USA\label{inst_stsci}
    \and Department of Physics and Astronomy, University of Heidelberg, Im Neuenheimer Feld 226, D-69120, Heidelberg, Germany\label{inst_hgsfp}}

   \date{Received 26 May 2026, Accepted 27 July 2026}

\abstract{
The SDSS-V Local Volume Mapper (LVM) is a wide-field integral field spectroscopic survey of the Southern Milky Way plane,
 the Magellanic Clouds, and nearby Local Group galaxies. We use Early Science observations of the whole body of the nearest nuclear cluster, 
Omega Centauri, to extend the LVM beyond its primary interstellar-medium science case.
The wide LVM field allows us to precisely map $\omega$\,Cen's line-of-sight rotation out to $\sim 3r_{HL}$ or $15^\prime$, 
reaching a maximum value of $(8.4 \pm 0.8)$\,km\,s$^{-1}$ at $r \approx 4.7^\prime$. Within the central region, comparisons 
with existing VLT MUSE oMEGACat data show explicitly that the unresolved-light signal is dominated by a small number of bright stars, 
with an effective sample size of only $\sim$12 per resolution element. Using Gaia DR3 as an external reference, 
we verify that the SDSS-V's LVM reduction pipeline recovers integrated stellar fluxes to 1--4\% across six magnitudes of surface brightness. 
Our  resulting total spectrum of $\omega$\,Cen is one of the highest S/N integrated spectrum for any globular or nuclear star cluster. We use it to test four widely-used SSP template libraries against resolved age-metallicity ground truth from oMEGACat. 
All templates recover an old, metal-poor population. But, even at S/N\,$\sim$1300, the inferred mean ages and mean [Fe/H] vary 
by $\sim$7\,Gyr and $\sim$0.4\,dex, respectively, across libraries and wavelength ranges, reflecting a systematic floor for 
integrated-light studies of old multi-population systems.}


   \maketitle
\nolinenumbers 
\section{Introduction}
The Local Volume Mapper (LVM) project \citep{2024AJ....168..198D} is one of the three programs that form the Sloan Digital Sky Survey-V \citep{2026AJ....171...52K}.
It targets the southern Milky Way plane, the Magellanic Clouds, nearby star-forming regions such as Orion \citep{2024A&A...689A.352K} and the Lagoon nebula \citep{2026ApJ..1001..238S} and the Rosette nebula \citep{2025MNRAS.543.1196V} and a sample of galaxies within the Local Volume using wide-field integral field spectroscopy (IFS). The scientific goal is to study the interstellar medium and resolve feedback mechanisms at parsec scales around individual sources. While the main targets of the LVM are the gas emission lines of the interstellar medium, in this paper, we study the unresolved stellar light of the ancient star cluster Omega Centauri (NGC 5139, hereafter \omc). This deliberately extends the use of the LVM beyond its primary interstellar-medium science case and explores its potential for stellar-continuum and integrated-light applications. This work serves as a demonstration for stellar population studies with the LVM, as a benchmark for extragalactic unresolved light studies, and finally yields precise measurements of \omc's line-of-sight rotation in its outer regions, where spatially complete spectroscopic coverage remains more limited than in the well-studied center.

In this introduction we first give a brief overview of unresolved studies of stellar populations, then we introduce \omc{} as an ideal benchmarking target and finally give an outline of the different sections of this paper.

In most galaxies outside the Local Group, the spatial resolution of current astronomical instruments is not sufficient to resolve individual stars. The observed stellar light therefore consists of the combined spectra of many individual stars, often overlaid with other radiating components such as interstellar gas, dust, and the emission of active galactic nuclei. The unresolved light is typically studied using stellar population synthesis \citep[e.g.][]{1976ApJ...203...52T,2003MNRAS.344.1000B}. A comprehensive review of the population synthesis method is given by \cite{2013ARA&A..51..393C}. Besides the stellar populations themselves, the shift and shape of absorption lines can also reveal the kinematics of the observed stellar population. The combination of chemical and kinematic properties of galaxies allows the study of their assembly histories, as reviewed by  \cite{2025ARA&A..63..259V}.
Important examples for surveys targeting large samples of galaxies using IFS are, e.g. the SAURON project \citep{2002MNRAS.329..513D} ,CALIFA \citep{2012A&A...538A...8S}, SDSS-IV MaNGA \citep{2015ApJ...798....7B} or (at higher spatial resolution and in the local universe) PHANGS-MUSE \citep{2022A&A...659A.191E} and GECKOS \citep{2024IAUS..377...27V,2025A&A...700A.237F}; see also \cite{2020ARA&A..58...99S}.

The study of integrated, unresolved light of individual star clusters is also of interest: the globular cluster population of a galaxy correlates with various galaxy properties \citep[e.g.][]{2013ApJ...772...82H} and can trace both in-situ and accreted components \citep[e.g.][]{2013MNRAS.436..122L,2019A&A...630L...4M}. Spectroscopic observations allow us to overcome the age-metallicity degeneracy \citep{1994ApJS...95..107W} of photometric observations, and at the same time, the line-of-sight (LOS) kinematics of globular clusters can be used to study the mass distribution in the halos of their galaxies \citep[e.g.][]{2024A&A...685A.132D}.
Some other recent works conducted with the VLT MUSE\footnote{Very Large Telescope Multi Unit Spectroscopic Explorer (VLT MUSE), \cite{2010SPIE.7735E..08B}} integral field spectrograph include the study of globular cluster populations in galaxy cluster environments such as Fornax \citep{2020A&A...637A..26F} and Hydra I \citep{2024A&A...683A...8G,2026A&A...705A.117M}, as well as around individual galaxies such as M104 \citep{2026OJAp....955146F}. An example of resolved kinematic measurements using IFS observations of unresolved light is the study by \cite{2022ApJ...924...48P} targeting M31's most massive globular cluster B023-G078.

The globular clusters of the Milky Way play a special role in this framework: On the one hand, they are important tracers of the Milky Way's accretion history \citep[e.g.][]{2019A&A...630L...4M}. On the other hand, they can be resolved into individual stars and therefore serve as benchmark targets to test and calibrate the methods used for unresolved stellar population studies. This has been applied e.g., by \cite{2002A&A...395...45P} or \cite{2020ApJ...896...13B}. Comprehensive surveys of the unresolved light of a large number of Milky Way globular clusters have been conducted by \cite{2005ApJS..160..163S} (40 clusters) and \cite{2017MNRAS.468.3828U} (WAGGS project, 64 clusters). 

\omc{} is the most massive star cluster in the halo of the Milky Way \citep{2018MNRAS.478.1520B}. Due to its proximity ($d=5.494$\,kpc, \citealt{2025ApJ...983...95H}) and its high surface brightness, it is arguably one of the best studied stellar systems and has revealed many complex and unusual properties. In comparison with typical globular clusters, it shows a large spread in age \citep[e.g.][]{2004A&A...422L...9H, 2007ApJ...663..296V, 2024ApJ...977...14C} and metallicity \citep[e.g.] []{1975ApJ...201L..71F, 2010ApJ...722.1373J,2021MNRAS.505.1645M,2024ApJ...970..152N}. These complexities have long motivated the scenario in which \omc{} is the stripped nucleus of a dwarf galaxy that has been accreted by the Milky Way billions of years ago \citep[e.g.][]{1999Natur.402...55L,2003MNRAS.346L..11B,2021MNRAS.500.2514P,2022ApJ...935..109L,2026MNRAS.547ag433M,2026arXiv260323589S}.

\omc's kinematics have also been extensively studied. Observations of the stellar proper motions and LOS velocities of individual stars revealed significant rotation \citep[e.g.][]{1986A&A...166..122M, 2018MNRAS.473.5591K,2024MNRAS.528.4941P,2024A&A...688A..92P,2024ApJ...970..192H}, an anisotropic velocity dispersion profile \citep{2015ApJ...803...29W, 2025ApJ...983...95H}, and, in combination with dynamical models, a relatively high mass-to-light ratio \citep[e.g.][]{2006A&A...445..513V, 2010ApJ...710.1063V,2013MNRAS.429.1887D, 2013MNRAS.436.2598W, 2022MNRAS.511.4251E}. The presence of fast-moving stars in the centermost arcseconds indicates the presence of an intermediate-mass black hole  \citep{2024Natur.631..285H}.

The most comprehensive observational dataset of the central regions of \omc{} has been the recent oMEGACat survey\footnote{oMEGACat project website: \url{https://omegacatalog.github.io/}} (PIs: Nadine Neumayer, Anil Seth), which combines a $\sim100$ pointing VLT MUSE mosaic \citep{2023ApJ...958....8N} with an extensive photometric and proper motion study based on almost a thousand individual Hubble Space Telescope (HST) observations \citep{2024ApJ...970..192H}. The oMEGACat catalog contains spectroscopic metallicity and LOS velocity measurements for around 300,000 resolved individual stars and high-precision astrometry and photometry for around 1.4 million stars within the half-light radius ($r_{\rm HL}=5\arcmin$; \citealt{2010arXiv1012.3224H}).

The availability of these resolved datasets and the complex stellar population properties of \omc{} make it an ideal benchmarking target for stellar population studies with the LVM and unresolved studies of stellar populations in general. The LVM spatial scale of $35\farcs3$/spaxel corresponds to around 1\,pc at the distance of \omc, comparable to what can be achieved with, for example, VLT MUSE wide-field mode (spaxel size: $0\farcs2$) at the distance of 1\,Mpc or with the narrow-field mode (spaxel size: $0\farcs025$) at distances of up to 8\,Mpc. Finally, the ability of the LVM to collect integrated stellar light over large contiguous regions allows us to study the bulk rotation in the outskirts of \omc, where other studies are limited to picking individual stars with multi-object spectrographs.
\autoref{sec:obs} presents a brief introduction about the LVM instruments and the observations of \omc. \autoref{sec:data_red} describes the data reduction process, including methods to identify foreground contamination, spatially bin the data, and the applied full spectrum fitting technique. \autoref{sec:kinematics} details the derivation of the two-dimensional kinematic maps, along with the corresponding corrections used to determine the rotation curve and its comparison with data from the MUSE spectrograph. \autoref{sec:stellar_pops} describes the study of \omc{}'s stellar populations and comparisons with resolved studies, and finally, \autoref{sec:summary} presents a summary and the conclusions of the work.

\section{Observations}
\label{sec:obs}
To achieve the LVM survey's science goals, an entirely new facility has been constructed -- the LVM Instrument \citep[LVM-I;][Blanc et al. in prep.]{2024SPIE13096E..1ZK}, which comprises three DESI-like spectrographs \citep{2018SPIE10702E..7KP,2022AJ....164..207D} that are fed by four 16-cm aperture telescopes in a siderostat configuration \citep{2024AJ....168..267H}. One telescope is dedicated to the science observations and features a hexagonal, fiber-based integral field unit (IFU) with 1801 fibers. The other three telescopes serve calibration purposes. Two contain a smaller IFU for measurements of the sky background, and one is equipped with a fiber-selector mechanism for observations of standard stars. Each fiber is equipped with microlenses and covers an on-sky area with a diameter $D=35\farcs3$. The fiber spacing is $37\arcsec$, yielding an overall filling factor of 83\%. The science IFU therefore has a large footprint with an approximate diameter of 0.5$\degr$. The spectrographs cover a large wavelength range (3600\,\AA–9800\,\AA ) at medium resolution ($R\sim4000$ at the wavelength of H$_\alpha$).

\begin{figure*}
    \centering
    \includegraphics[width=1.0\linewidth]{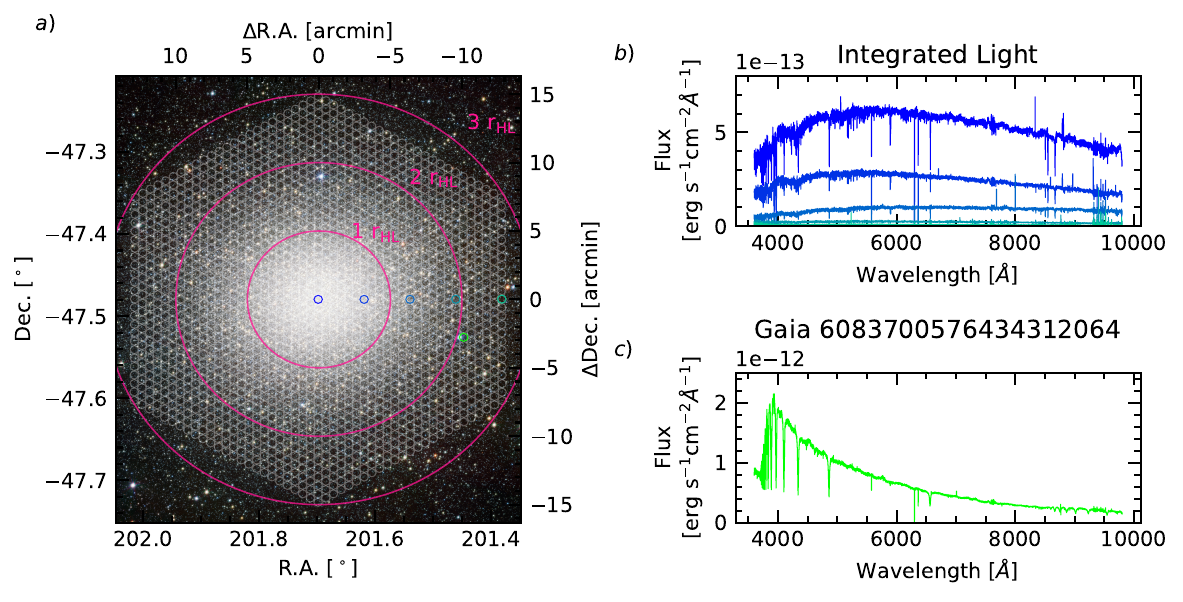}
     \caption{Overview of the LVM observations of \omc. \textit{a)} shows the footprint of the LVM Science IFU over a ground-based wide field image of \omc. \textit{b)} shows the spectra of five individual fibers at different cluster radii (indicated by blue/green circles in panel \textit{a)}), illustrating the decrease in surface brightness. \textit{c)} shows the spectrum of a fiber that is dominated by a single bright (Gaia $G = 8.8$) foreground star. This star is picked up by the fiber indicated by a bright green circle that falls next to the magenta large circle. The large magenta circles mark the 1, 2 and 3 half-light radius (r$_{\mathrm{HL}}$) of \omc{}.\\
     \textbf{Image Credit:} ESO/INAF-VST/OmegaCAM. Acknowledgement: A. Grado, L. Limatola/INAF-Capodimonte Observatory}
      \label{fig:overview}
\end{figure*}

\omc{} was chosen as an Early Science Target for the LVM, and at the time of writing, four individual exposures of \omc{} have been observed. \autoref{fig:overview} \textit{a)} shows the footprint of the individual LVM fibers on a ground-based wide field image of \omc{}. A single LVM pointing covers around 3 half-light radii ($3\,r_{\rm HL}=15\arcmin$) of \omc{}.
The first exposure (LVM Exposure ID: 3499) was taken during the instrument commissioning in July 2023 and lacks essential calibration data; therefore, we only use the three later exposures (LVM Exposure  ID: 20836, 20837, 20838) that were taken in July 2024. All exposures have the LVM survey standard exposure time of 900\,s. There is no pointing-offset between these three exposures, meaning that gaps between fibers remain uncovered. On the other hand, this allows for easy stacking of the spectra for individual fibers.

\section{Data Reduction}
\label{sec:data_red}
\subsection{Data Reduction Pipeline}
The raw data were reduced using the standard LVM data reduction pipeline (Mejía-Narváez et al. in prep) in the internal release version \texttt{DRP v1.2.1}. With respect to previous versions, this version includes various improvements to the flux and wavelength calibrations and the implementation of telluric corrections.

\subsection{Verification of the LVM flux calibration}
\begin{figure*}[h]
    \centering
    \includegraphics[width=1.0\linewidth]{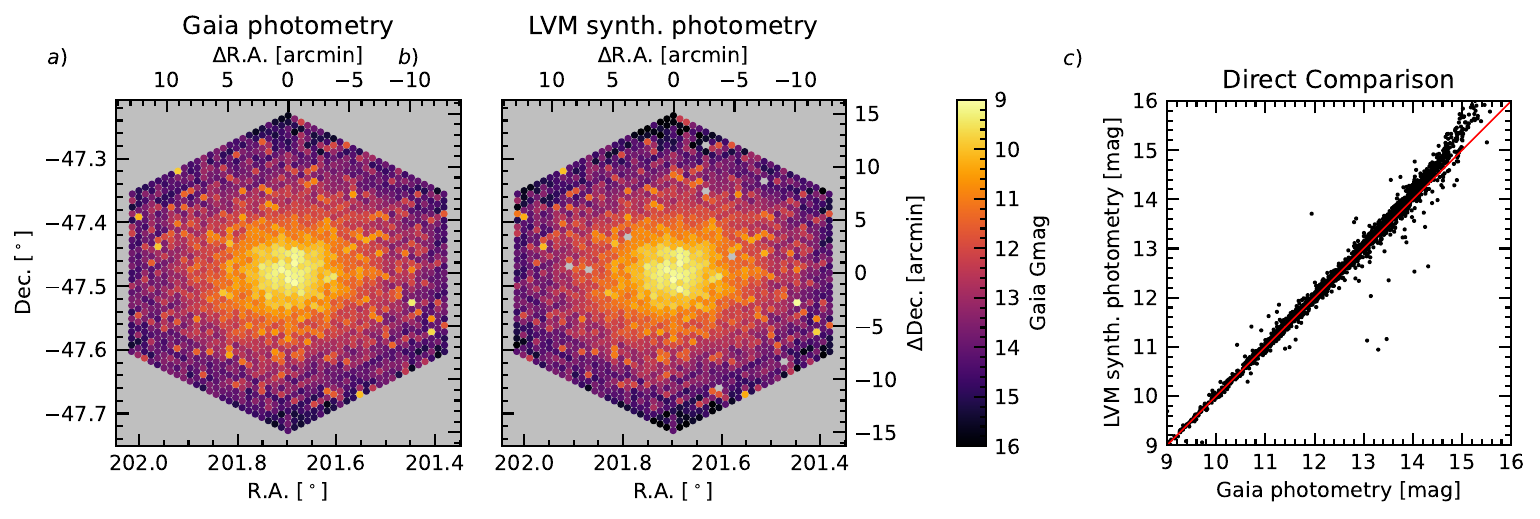}
     \caption{Illustration of the flux calibration verification. \textit{a)} shows the total integrated Gaia G magnitude of all Gaia DR3 stars contained in each LVM fiber. \textit{b)} shows the synthetic Gaia G magnitude as calculated from the individual spectra measured with the LVM. \textit{c)} shows a direct comparison between the magnitude values from \textit{a)} and \textit{b)}.}
      \label{fig:fluxcalibration}
\end{figure*}

The \omc{} observations provide a unique opportunity to verify the accuracy of the LVM flux calibration for continuum emission over a wide dynamic range. As a reference, we used the space-based catalog of the third Gaia data release  \citep{2023A&A...674A...1G}. Due to the limited completeness of Gaia DR3 in the crowded center of \omc{} we complemented the nominal Gaia catalog with the dedicated Focussed Product Release on Omega Centauri \citep{2023A&A...680A..35G}.\footnote{See Appendix B of \cite{2023A&A...680A..35G} for the query used to download the combined Gaia DR3+FPR catalog.}. We then determined all stars that would fall on each fiber, assuming the nominal fiber diameter of 35$\farcs$3, and calculated their combined magnitude in the Gaia G band. The number of Gaia stars per fiber ranged between 23 in the cluster outskirts and 2321 in the central region. We then convolved the individual calibrated LVM spectra with the Gaia passband and calculated synthetic magnitudes using the \texttt{pyphot} package \citep{2025zndo..14712174F} which we then compared with the space-based Gaia value. \autoref{fig:fluxcalibration} shows both the resulting integrated flux maps and the comparison plot. The agreement between the two is excellent: we find a median offset of 0.06\,mag for the bright spaxels (with an integrated G mag < 12) and a median offset of 0.09\,mag over all spaxels. This further decreases to 0.01\,mag (bright spaxels only) and 0.043\,mag (all fibers) when using a smooth function (see Apendix A) to attenuate the flux of stars at edge of the spaxels. This corresponds to an absolute flux uncertainty of 1\% and 4\%, meeting the LVM requirement of 10\% \citep{2024AJ....168..198D} and matching other tests of the LVM flux calibration \citep{2026ApJ...997..339H}. The RMS of the difference between Gaia and synthetic magnitudes is 0.11\,mag after 3$\sigma$ clipping. Individual strong outliers (with a maximum of 2.5 mag) can be attributed to individual bright stars that fall on the edge of a spaxel, however, only 16 fibers ($<$1\% of all fibers) show a deviation larger than one magnitude. 

These flux calibration experiments also allow for a verification of the astrometric accuracy of the LVM data products. We systematically shifted the nominal right ascension (R.A.) and declination (Dec.) of each fiber using a grid with 0.25\arcsec\ spacing and a range of $\pm10$\arcsec\, and then recalculated the integrated Gaia fluxes. The best agreement was achieved at an offset of $\Delta\rm R.A.=2.5\arcsec$ and $\Delta\rm Dec.=0\arcsec$. Given the fiber-size of 35$\farcs$3 and the nominal point-spread-function size of FWHM$<3.5 \arcsec$, this is a good result and confirms the laboratory focal-plane metrology \citep{2022SPIE12184E..6UH, 2024AJ....168..267H}.

\subsection{Identification of foreground contamination}
\label{sec:contamination}
\begin{figure*}
    \centering
    \includegraphics[width=1.0\linewidth]{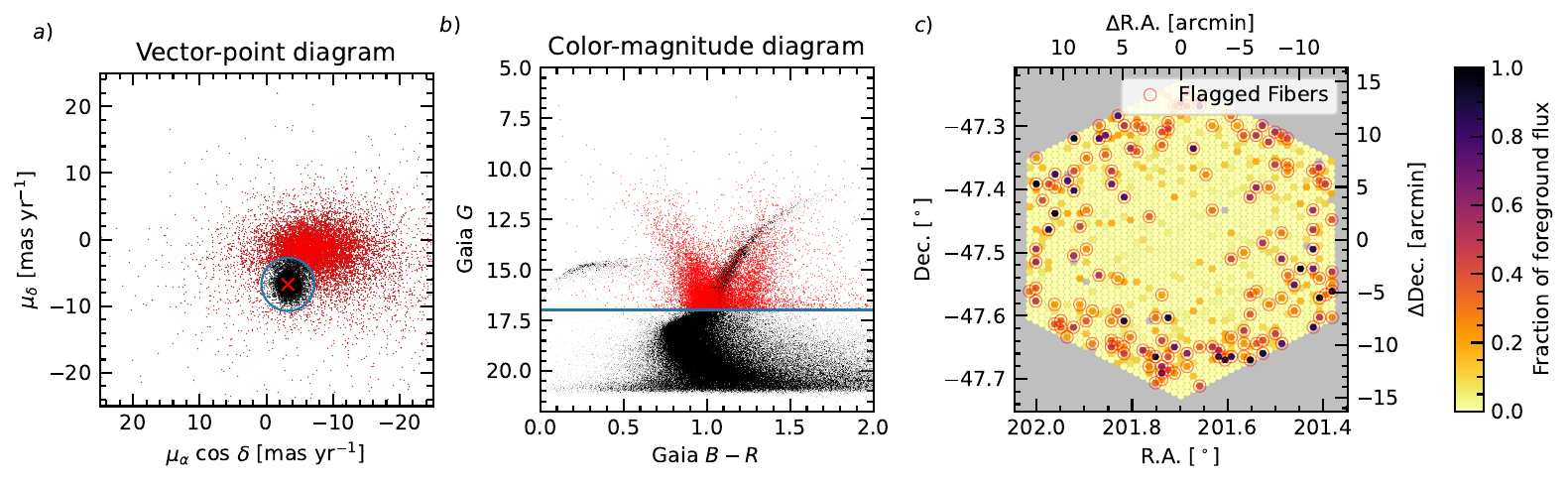}
     \caption{Illustration of the identification of contaminated fibers. \textit{a)} shows the two proper-motions of all Gaia DR3 sources brighter than $G=17$ within 0.8$^\circ$ of \omc's center. Stars with proper motions larger than a 4\,mas\,yr$^{-1}$ with respect to the mean-cluster motion (red) are considered as contaminants. \textit{b)} demonstrates the effect of the astrometric selection in a Gaia-based color-magnitude diagram. While the cluster member stars (black) lie on the expected sequences for a globular cluster CMD, the foreground stars (red) show a very different distribution. \textit{c)} shows the flux contribution of all foreground stars to the total flux in each fiber. While in the center, contamination is typically close to negligible, in the outskirts, several fibers are dominated by the foreground. We flag all fibers with a contamination fraction\,\,$>0.2$.}
      \label{fig:foreground}
\end{figure*}

Although \omc{} is located above the Galactic plane, there is a non-negligible number of foreground stars that contaminate the unresolved light spectrum of the cluster. An example is shown in \autoref{fig:overview} \textit{c)}, where a spaxel is dominated by the light of a single Gaia $G = 8.8$, A-type star. To quantify the level of contamination and flag heavily contaminated fibers, we considered all sources from the Gaia DR3+FPR catalog \citep{2023A&A...674A...1G} within a 0.8\degr\ radius of \omc's center. To have a high-quality sample, we then only selected sources brighter than Gaia $G = 17.0$. 
In the next step we selected all stars with a proper motion that differs by at least 4\,mas\,yr$^{-1}$ (5 times the central velocity dispersion) from the mean motion of \omc{} ($(\mu_\alpha\cos\delta, \mu_\delta)=(-3.238,-6.716)\,$mas\,yr$^{-1}$ as derived by \citealt{2018ApJ...854...45L}). Panel $a)$ in \autoref{fig:foreground} shows this astrometric selection. Panel $b)$ demonstrates how this effectively separates the foreground and the cluster member stars in a Gaia-based color-magnitude diagram. Finally, we collected all Gaia stars that fall into a specific fiber and determined the contribution of foreground stars to the total flux of each fiber using the Gaia $G$ band magnitudes. Panel $c)$ in \autoref{fig:foreground} shows this ratio for all fibers. It is evident that in the outskirts of \omc{} many fibers are significantly contaminated, while in the center the contamination fraction is close to 0, meaning that the contribution of foreground stars is negligible. For the rest of the analysis, we exclude all fibers with at least 20\% foreground contamination. We note that the magnitude cut in this method certainly ignores a number of fainter foreground stars. However, these stars have a low relative contribution to the total flux, even the faintest fibers in the outskirts have an integrated magnitude of at least Gaia $G \leq 15$. In addition, a small fraction of foreground stars with proper motions close to the systemic proper motion of \omc{} is not identified, however we deem this residual contamination acceptable  - a perfect foreground separation would never be possible anyway due to the large on-sky diameter of the LVM fibers.

\subsection{Spatial binning and normalization}
\label{subsec:binning}
As there was no significant pointing offset between the three exposures, we co-added the spectra for each fiber to increase the signal-to-noise ratio (S/N) by a factor $\sqrt{3}\approx 1.73$. We also added the nominal variances for each spectral value. This step indeed improved the results of the various full-spectrum fits described in the next sections (compared to using individual exposures), which is why we adopted it as standard for this analysis.

In the center of \omc{}, the S/N of individual fibers is high and reaches a value of up to 148 per fiber (measured as median over the full spectral range and using the 3 co-added exposures). In the outskirts, it is significantly lower, preventing successful full-spectrum fits. Exemplary spectra for individual fibers at different radii are shown in \autoref{fig:overview}, \textit{b)}. To tackle this, we combined the light from different spectra and used the Voronoi binning scheme \citep{2003MNRAS.342..345C} with a target S/N of 75 per bin. This provides the best compromise between stable results of the kinematic fits and spatial resolution. As the target S/N is lower than the S/N in the brightest central spaxels, typically no binning occurs within the half-light radius, while in the faint outskirts, up to 79 individual fibers are binned. The total number of individual bins is 299.

For each Voronoi bin, we calculate a combined, normalized spectrum by first dividing each individual fiber's spectrum by its median and then calculating the 3$\sigma$-clipped mean for each spectral element across all fibers within the bin. An alternative approach (which we use for the creation of very high-S/N spectra for stellar population studies) would be to simply add them, however, in that case, individual fibers with a single bright star could dominate their whole bin. The normalization-based combination therefore reduces the influence of stochastic bright-star contributions when the goal is to measure the average kinematic behavior of the unresolved stellar population.

In addition to the Voronoi binning scheme, we calculate the sum of all spectra from uncontaminated fibers within the half-light radius. The goal of this step is to create a very high S/N spectrum that can be used to study the stellar population properties of \omc{} (see \autoref{sec:stellar_pops}). Unlike the normalized Voronoi spectra, this summed spectrum preserves the luminosity-weighted integrated light within the half-light radius and is therefore better suited for stellar-population measurements that rely on the relative contribution of different stellar components to the total flux.

\subsection{Full Spectrum Fitting}
To derive the mean velocity and stellar population information, we use the full spectral fitting code \texttt{pPXF} \citep{2004PASP..116..138C, 2017MNRAS.466..798C,2023MNRAS.526.3273C}. This code fits spectra by determining the optimal linear combination of various single stellar population (SSP) spectral templates. To account for the kinematics, these spectra are shifted and convolved with Gauss-Hermite functions with a chosen degree. The detailed parameters for the different fits are described in the respective sections (\ref{sec:kinematics} and \ref{sec:stellar_pops}).

\section{Kinematics}
\label{sec:kinematics}
\subsection{Spatially Resolved Line-of-Sight Velocity Maps}

For our kinematic analysis, we fit the combined spectrum for each of the 299 Voronoi bins with the simple stellar population models from the XShooter Stellar Library \citep[XSL;][]{2022A&A...661A..50V}. The XSL SSP templates are a natural choice, as they cover the full LVM wavelength range with significantly higher resolution ($R\sim10\,000)$  than the observed spectra.

\begin{figure}
    \centering
    \includegraphics[width=1.0\linewidth]{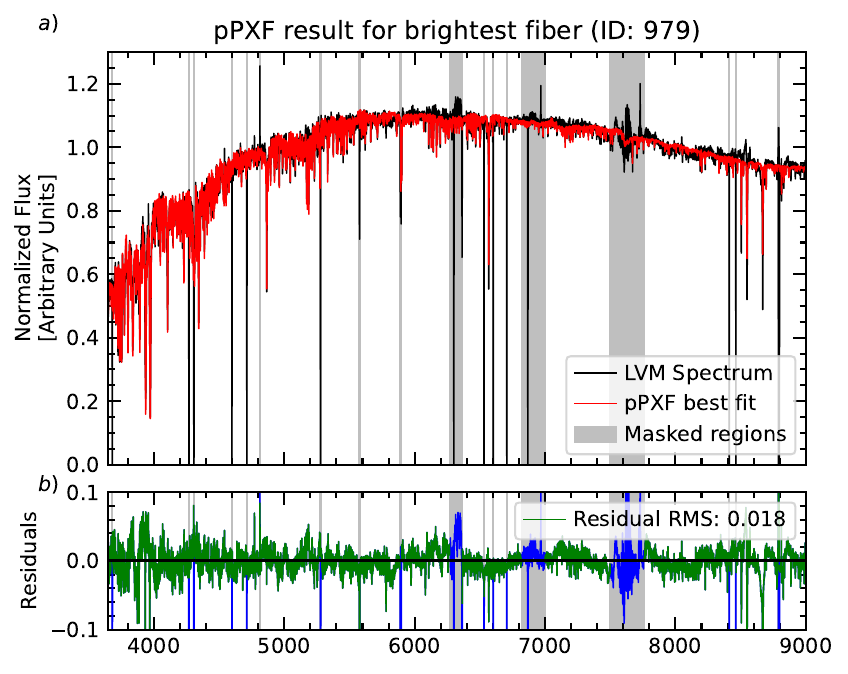}
     \caption{Result of the pPXF fit for a high S/N spectrum, based on a single fiber located $\sim$1.5\,arcmin South of \omc's center. Panel \textit{a)} shows the input spectrum (black) and the resulting best fit (red). Masked regions are shaded in gray. \textit{b)} shows the fit residuals, which have an RMS of 1.8\% (with masked regions shown in blue).}
      \label{fig:ppxf_fit}
\end{figure}

We restrict the LVM spectra to the range $3650\,\AA < \lambda<9000\,\AA$, to avoid the low S/N region at the blue end, and the heavily sky-contaminated region at very long wavelengths.
For the fit, we use additive polynomials of degree\footnote{A polynomial order larger than 10 did not lead to a further decrease in the fit residuals.} $n=10$ and only the first two Gauss Hermite moments (mean velocity and $\sigma$). In the initial fit, we masked the wavelength ranges from $5887-5899\AA, 6268-6363\AA, 6820-7000\AA$, and $7500-7760\AA$, as they either lie in the transition regions of two spectrograph channels or are contaminated by foreground extinction lines (see \autoref{subsec:extinction}) or residual telluric absorption and sky emission. To remove individual outliers, we repeated the fit, while also masking all spectral elements that had residuals larger than 0.2 (normalized flux). All fits converged well with typical residuals of 2\% (RMS). An example of the fit of the highest S/N spectrum is given in \autoref{fig:ppxf_fit}. To estimate the uncertainty on the mean LOS velocity, we performed a bootstrapping analysis, in which we randomly sampled from the residual distribution and added these values to the input spectrum of each fit. We repeated this step $N=100$ times and calculated the standard deviation of the resulting parameters. The uncertainty in the mean LOS velocity ranged from 0.4\,km\,s$^{-1}$ to 1.7\,km\,s$^{-1}$ for the different bins and showed a clear correlation with the S/N.

The typical resolution for the LVM spectra is ${R=\lambda/\Delta\lambda\approx4000}$, yielding a velocity resolution of about 75\,km\,s$^{-1}$. At the calcium triplet, the velocity resolution is $\sim$60\,km\,s$^{-1}$ (FWHM), corresponding to a dispersion of $\sigma_{\rm inst.}\approx25.5$\,km\,s$^{-1}$.  This is significantly larger than the velocity dispersion of \omc, which reaches a central maximum of about 20\,km\,s$^{-1}$ and decreases to about 13.5\,km\,s$^{-1}$ at the half-light radius \citep{2025ApJ...983...95H}. Therefore, we do not attempt to measure the velocity dispersion of \omc{} from the LVM spectra. Higher-order moments of the velocity distribution ($h_3, h_4$) are not considered either. Due to the overall Gaussian-like velocity distribution in \omc{} we do not expect that the missing higher order moments introduce significant biases in the inferred mean velocities.

\begin{figure*}
    \centering
    \includegraphics[width=1.0\linewidth]{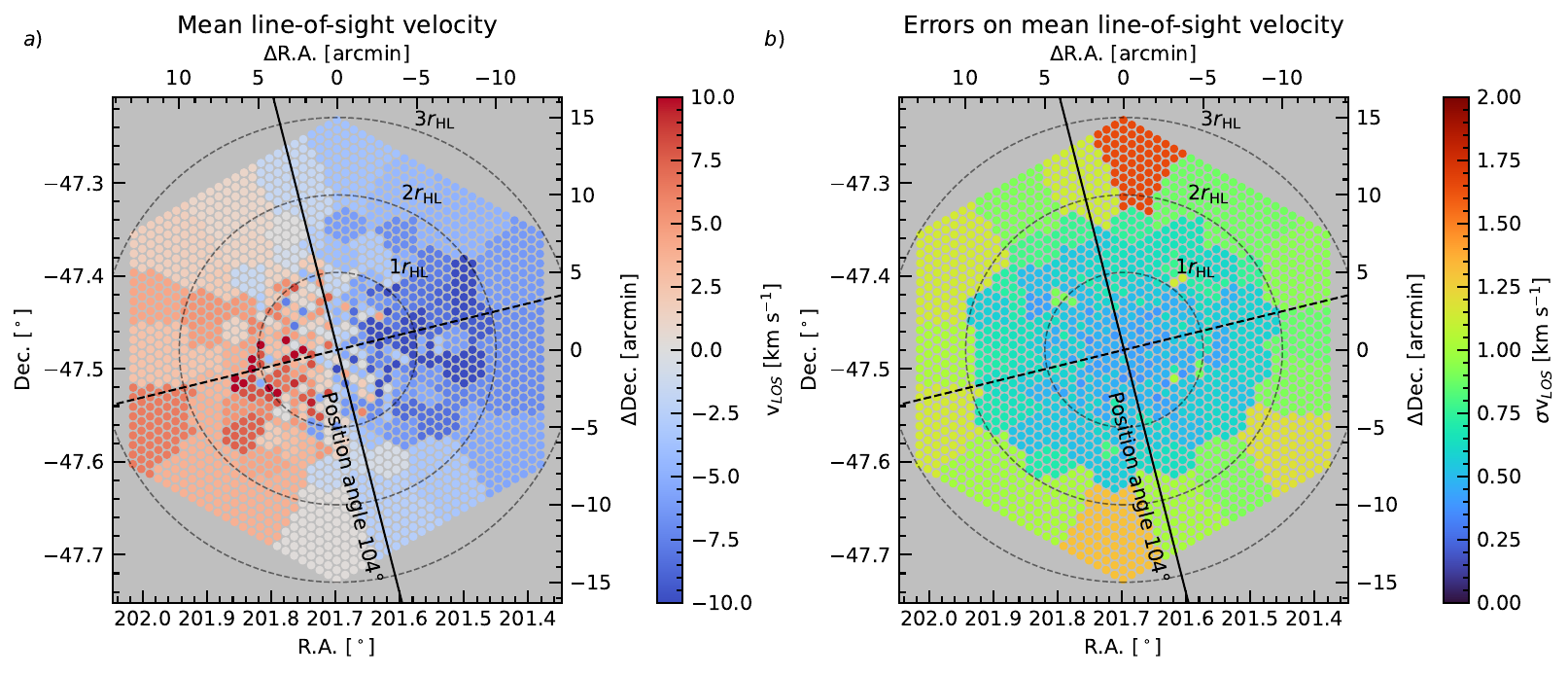}
     \caption{Relative LVM LOS velocity map and its uncertainty. \textit{a)} shows the derived mean LOS velocity corrected for the bulk motion of the cluster and for heliocentric motion. In the outskirts, typically many fibers are binned, while measurements within $r_{\rm HL}$ are based on individual fibers. The rotation of the cluster is clearly visible. We also show the position angle of the rotation axis as derived by \cite{2025ApJ...983...95H}. \textit{b)} shows the derived uncertainties on the LOS velocity measurements derived via bootstrapping. Uncertainties are higher at larger radii, where the S/N in the individual bins is lower.}
      \label{fig:lvm_losmap}
\end{figure*}

We show the results of our determination of the mean LOS velocity in \autoref{fig:lvm_losmap} \textit{a)} and the corresponding uncertainties in \textit{b)}. The map has been corrected for the bulk LOS velocity of the cluster (233\,km\,s$^{-1}$, \citealt{2023ApJ...958....8N}) and for the heliocentric velocity at the time of the observations (21.6\,km\,s$^{-1}$). The 2D velocity map shows a clear rotation pattern, which we study in more detail in \autoref{subsec:rotation}.

\subsection{Comparisons with resolved kinematic studies}

\begin{figure*}[h]
    \centering
    \includegraphics[width=0.85\linewidth]{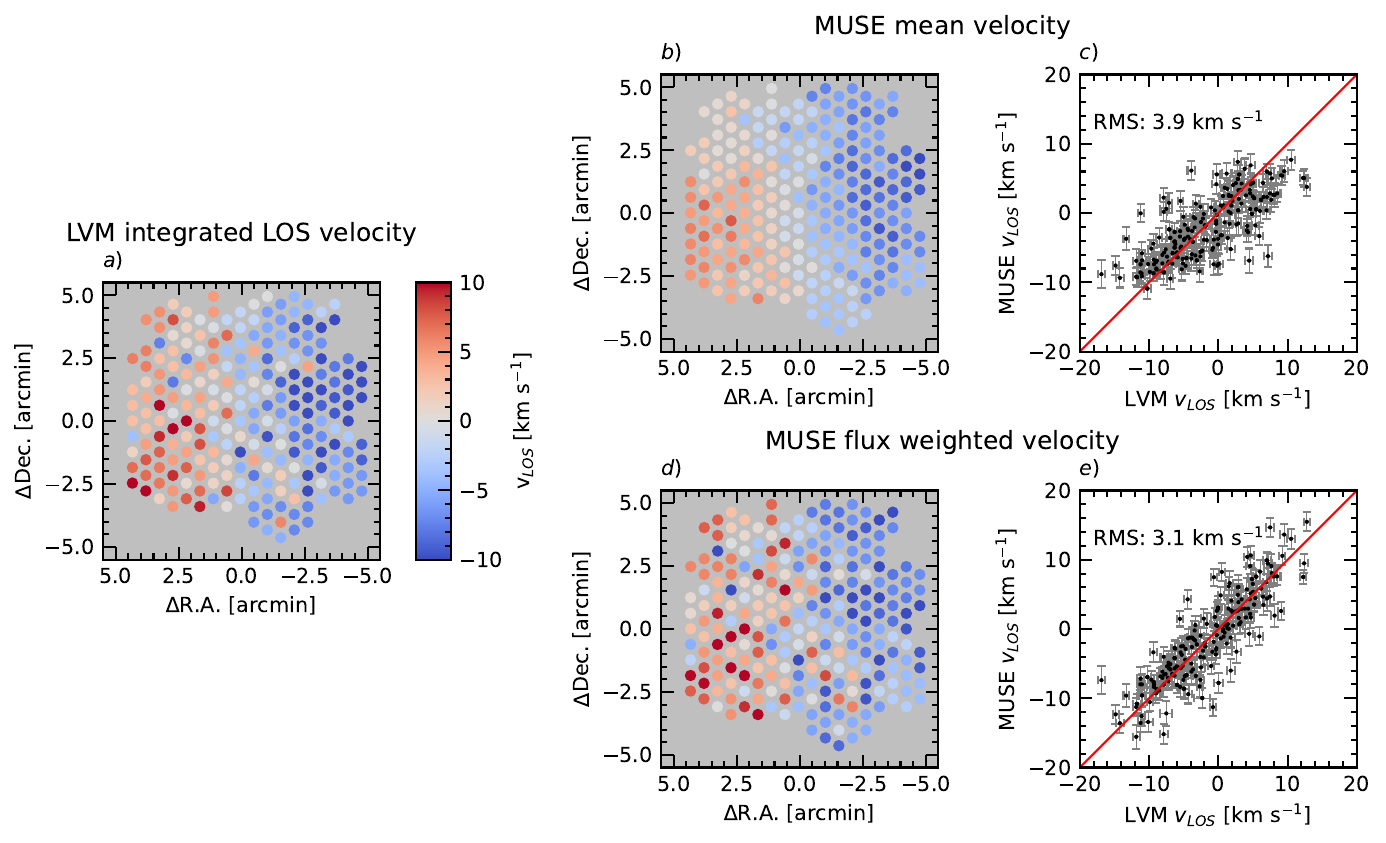}
     \caption{Comparison between the mean LOS velocities measured with the LVM using unresolved light and with VLT MUSE using resolved stars. \textit{a)} shows a map of the LVM LOS velocity measurements for the central region of \omc{} that overlaps with the MUSE coverage \textit{b)} shows a map of the (unweighted, arithmetic) mean of the individual stellar velocities in the MUSE sample using the same bins as for the LVM. \textit{c)} shows the direct comparison between the LVM measurements and the unweighted mean of the MUSE measurements. \textit{d)} shows a map of the mean of the MUSE measurements, weighted by the flux in the F625W filter. \textit{e)} shows the comparison between these flux-weighted mean values and the LVM data. The agreement when using the flux-weighted velocity is significantly better than with the unweighted mean. }
      \label{fig:muse}
\end{figure*}

Within the half-light radius of \omc{}, we can directly compare the unresolved-light LOS velocity results with the large number of individual stellar velocity measurements determined in the oMEGACat project \citep{2023ApJ...958....8N}. We used the public catalog of individual LOS velocities and made use of the LOS quality cuts described by \citealt{2025ApJ...983...95H} (\texttt{selection\_hq\_los==1}).
We then determined all well-measured stars that would fall into each LVM spaxel, and grouped them into the same bins as described in \autoref{subsec:binning}. The most populated LVM bin contains 291 MUSE stars, and we set a lower limit of 50 MUSE stars for the comparison, to reject bins with only partial spatial overlap. For each bin, we then determined the mean LOS velocity, using the arithmetic mean. The results of this step are compared with the LVM LOS velocities in \autoref{fig:muse}, \textit{b) c)}. The results show an interesting effect: while the global rotation pattern is similar in both datasets, the MUSE map is significantly smoother and shows less extreme values for individual spectra.  The reason for these deviations does not lie in uncertainties in the LVM spectral fits. Instead, it lies in the nature of the unresolved light measurements: while in the resolved MUSE data, all stars above the cut-off magnitude are attributed the same weight, the LVM spectra and the resulting fit velocities are dominated by a few bright stars. In the center of \omc{}, these stars have a velocity dispersion of up to 20\,km\,s$^{-1}$ and therefore add significant scatter on top of the mean rotation pattern.

To demonstrate this effect, we recalculated the MUSE LOS velocities for each bin, this time by weighting each individual stellar velocity value with the flux of the star, as measured with the HST F625W filter\footnote{We also tried bluer (F435W) and redder (F814W) filters available in the oMEGACat II for the flux-weighting, however, F625W provided the best agreement with the LVM data.}. The results for this flux-weighted averaging scheme are shown in \autoref{fig:muse}, \textit{d) e)}. The agreement with the LVM data is better both qualitatively (one can easily identify matching patterns in both the LVM and the MUSE maps) and quantitatively (the RMS of the difference between MUSE and LVM decreased from 3.9\,km\,s$^{-1}$ to  3.1\,km\,s$^{-1}$). The main reason for the remaining deviations even after applying the flux-weighting could be that the MUSE catalog is not fully complete and misses the light of all unevolved stars fainter than the turnoff. Furthermore, the wavelength range of the HST filters used to perform the flux weighting is small with respect to the wavelength range probed with the LVM.

We can also look at the flux-weighting effect by calculating the effective sample size $n_{\rm eff}=(\sum w_i)^2/\sum w_i^2$ \citep{Kish1995-pz} with the individual weights $w_i$ being the flux values of the stars. The effective sample size of the weighted mean gives the approximate sample size of an unweighted sample that would yield similar precision.

For our example, the median $n_{\rm eff}$ is 11.5, meaning that typically around 12 bright stars dominate the mean signal and scatter. This is significantly lower than the actual number of measured stars in each bin (median number of MUSE stars: 129). With a velocity dispersion of 13$-$20\,km\,s$^{-1}$ \citep{2025ApJ...983...95H}, this makes an additional scatter of our weighted mean values of 13$-$20\,km\,s$^{-1}/\sqrt{11.5}$=3.8$-$5.9\,km\,s$^{-1}$ plausible.

Finally, we note that the LVM LOS velocity map in the outskirts is much smoother than within the half-light radius for two combined reasons: First, the velocity dispersion decreases significantly at larger distances from the cluster center (\citealt{2018ApJ...853...86B} measure a velocity dispersion of 8\,km\,s$^{-1}$ at 3\,$r_{\rm HL}$). Second, in the outskirts, typically many fibers are combined and in our binning scheme we weight their contribution equally, which down-weights individual spaxels with bright stars.

\subsection{Perspective Rotation Corrections}
The space-motion of \omc{} causes an apparent rotation signal in the LOS direction, the so-called ``perspective rotation''. Following Eq. (5) in \cite{1997AJ....114.1074M} it can be calculated using
\begin{equation}
    v_{\rm pr}=\frac{\Delta x}{d}v_x+\frac{\Delta y}{d}v_y
\end{equation}
with $d$ being the distance to \omc,  $\Delta x$ and $\Delta y$ being the relative projected distance with respect to the cluster center in the two plane-of-sky directions, and $v_x$ and $v_y$ being the two velocity components. We assume $d=5.494$\,kpc and the absolute proper motion value of \cite{2018ApJ...854...45L} (see \autoref{sec:contamination}), which results in perspective rotation values ranging from $-0.62$ to $0.61$\,km\,s$^{-1}$. While this is smaller than the typical measurement uncertainty for individual bins, it should be taken into account when studying the overall rotation structure of \omc.
\subsection{Determination of \omc's rotation-curve}
\label{subsec:rotation}

\begin{figure}
    \centering
    \includegraphics[width=1.0\linewidth]{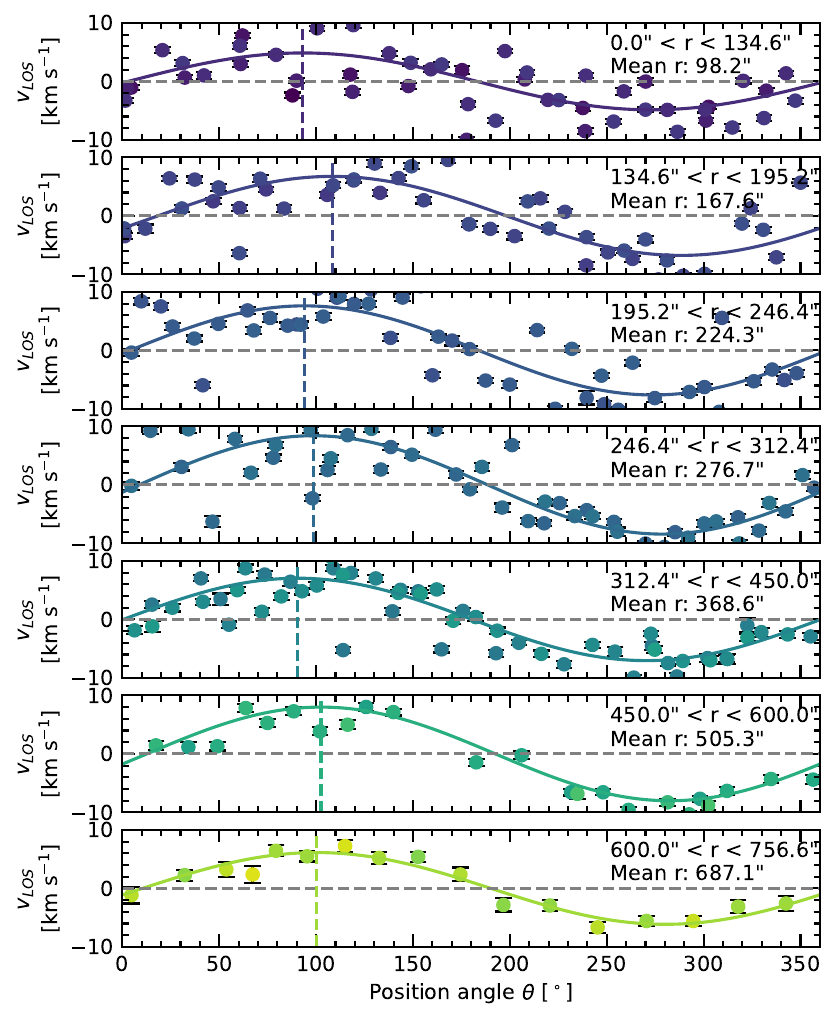}
     \caption{Determination of the rotation curve. In each panel, we plot the velocity measurements against their position angle for different radial bins. The different radial bins are also indicated with different colors. The best-fit sinusoidal function provides the rotation amplitude and the position angle at different radii.}
      \label{fig:rotation_fit}
\end{figure}

\begin{figure}
    \centering
    \includegraphics[width=1.0\linewidth]{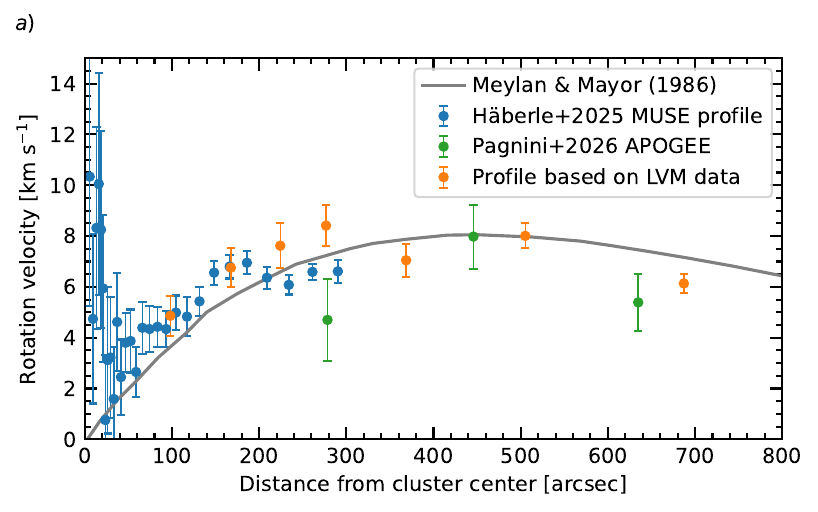}
    \includegraphics[width=0.66\linewidth]{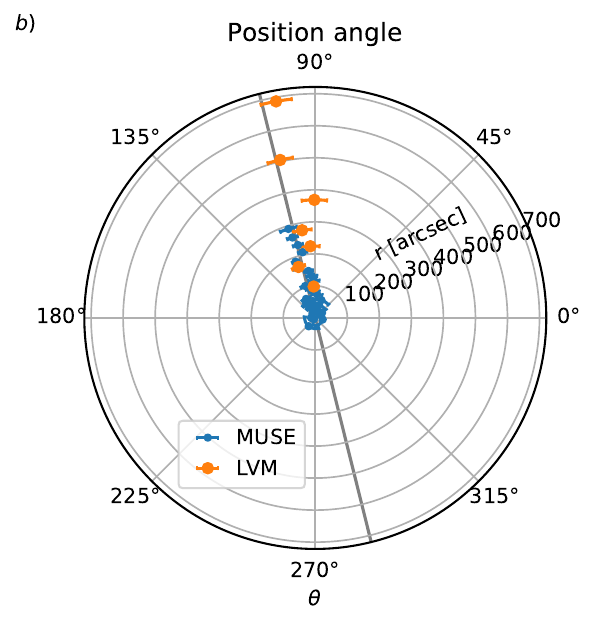}
     \caption{Result for the rotation curve and position angle. \textit{a)} shows the rotation profile measured with the LVM in comparison with the resolved VLT MUSE results of \cite{2025ApJ...983...95H}, \textit{b)} shows the position angle. The gray line indicates a position angle of 104$^\circ$, the best fit value derived from the MUSE data.}
      \label{fig:rotation_results}
\end{figure}

To estimate the LOS velocity rotation curve for \omc{} based on our LVM velocity map, we first calculate the position angle with respect to the \cite{2010ApJ...710.1032A} cluster center for each of our 299 Voronoi bins. We then distribute the data into 7 approximately equally populated radial bins (the five inner radial bins contain 50 values, the two outermost bins 32 and 18 values; this provided the best compromise between spatial resolution and SNR) and study how the mean LOS velocity (corrected for perspective rotation) varies with the position angle. We then fit the relation between velocity and position angle with:
\begin{equation}
    v_{\rm LOS}(\theta)=A_{\rm rot}\cos{(\theta -\theta_0)},
\end{equation}
where the two free parameters are the rotation amplitude $A_{\rm rot}$ and the position angle of the rotation axis $\theta_0$. An illustration of this analysis step is shown in \autoref{fig:rotation_fit}, and the resulting rotation profile is shown in \autoref{fig:rotation_results}. The numerical values of the rotation profile are given in \autoref{tab:rotation_profile} and shared with the electronic material accompanying the paper.

\begin{table}[]
\centering
\caption{Numerical values of \omc's rotation profile. }
\label{tab:rotation_profile}
\begin{tabular}{llllll}
\hline
r$_{\rm min.}$ & r$_{\rm median}$ & r$_{\rm max.}$ & $v_{\rm rot., LOS}$ & P.A. $\theta$ & $N$ \\ \relax
{[}$\arcsec${]}   & {[}$\arcsec${]}     & {[}$\arcsec${]}   & {[}km\,s$^{-1}${]} & {[}degree{]}  &   \\ \hline
  0.0 &  98.2 & 134.6 &   4.86 $\pm$0.78 &   93.0 $\pm$9.2 & 50 \\
134.6 & 167.6 & 195.2 &   6.76 $\pm$0.78 &  108.3 $\pm$7.3 & 49 \\
195.2 & 224.3 & 246.4 &   7.62 $\pm$0.89 &   93.9 $\pm$7.4 & 49 \\
246.4 & 276.7 & 312.4 &   8.41 $\pm$0.81 &   98.6 $\pm$6.2 & 49 \\
312.4 & 368.6 & 450.0 &   7.05 $\pm$0.66 &   90.5 $\pm$6.2 & 58 \\
450.0 & 505.3 & 600.0 &   8.01 $\pm$0.49 &  102.5 $\pm$4.6 & 23 \\
600.0 & 687.1 & 756.6 &   6.13 $\pm$0.38 &  100.3 $\pm$4.1 & 17 \\ \hline
\end{tabular}
\end{table}

We observe a maximum rotation value of (8.41$\pm$0.81)\,km\,s$^{-1}$ at 276\arcsec\ that slowly decreases towards larger radii. Within the half-light radius, both the rotation amplitude and the rotation curve are in agreement with the results of \cite{2025ApJ...983...95H}, but the LVM data extend to 2.5 times larger radii. At these large radii, suitable comparison datasets are rare.

The seminal work of \cite{1986A&A...166..122M} assumed a functional form for the rotation profile, and obtained a peak velocity of 8.0\,km\,s$^{-1}$ at $7\farcm2$ ($=475\arcsec$). Most recently, \cite{2026arXiv260520503P} used APOGEE spectra to determine the rotation in \omc's outskirts. We show the three overlapping datapoints in \autoref{fig:rotation_results}. Due to the relatively low number of individual stellar velocities, the statistical errors of this dataset are significantly larger than on our measurements. Overall we see good agreement, however, the innermost datapoint from the \cite{2026arXiv260520503P} analysis is between 1-2$\sigma$ lower than our measurements.
Other works use astrometric proper motion measurements to determine the rotation curve in the plane of the sky \citep{2013ApJ...772...67B,2018MNRAS.481.2125B, 2021MNRAS.505.5978V,2023MNRAS.522L..61S}. Taking into account the distance to \omc{} and its inclination, these profiles qualitatively agree with our new LOS rotation curve.

\subsection{A combined MUSE LVM line-of-sight velocity map}
Within the half-light radius, the resolved VLT MUSE data from oMEGACat provide better constraints for dynamical models due to the higher spatial resolution and the consideration of a significantly larger number of stars (see discussion in the previous section). The MUSE measurements also allow us to determine the LOS velocity dispersion. However, the LVM data cover a significantly larger spatial extent and reach out to up to three times the half-light radius.

To facilitate future dynamical modeling efforts, we construct and make publicly available a combined mean LOS velocity map that contains the MUSE mean LOS velocity measurements in the center and the new LVM measurements in the outskirts. This combined map contains all MUSE data points published by \cite{2025ApJ...983...95H}, and all LVM bins from this work, whose centers are at least $35\farcs3$ (i.e. one LVM fiber diameter) away from the center of a MUSE bin. The resulting combined map is shown in \autoref{fig:combined_map}. We describe the published table in \autoref{appendix:published_table} and show exemplary table rows in \autoref{tab:combined_map}.
\begin{figure}
    \centering
    \includegraphics[width=1.0\linewidth]{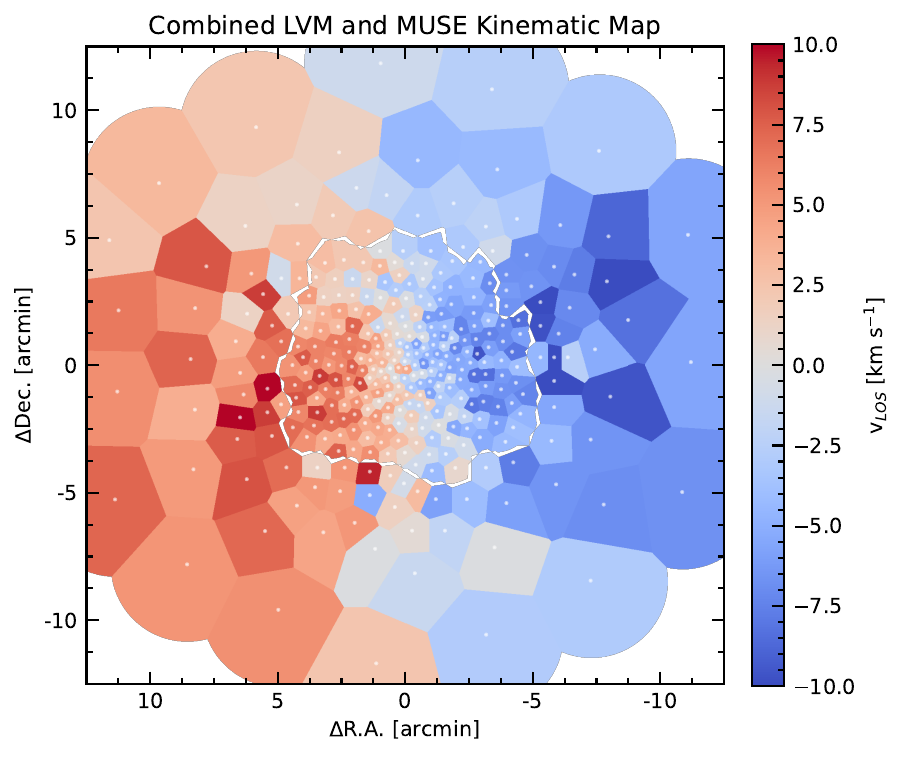}
     \caption{Combined LOS velocity map. This figure shows a combined LOS velocity map of \omc. The inner region (separated by a white border) is based on individual VLT MUSE measurements and taken from \cite{2025ApJ...983...95H}, the outer part is based on the unresolved LVM measurements described in this work. The bin centers are marked with white dots, the map is created using nearest-neighbor interpolation.}
      \label{fig:combined_map}
\end{figure}

\section{Stellar Populations}
\label{sec:stellar_pops}
\subsection{Resolved reference data}
The most detailed study of the age-metallicity relation of \omc{} is the recent work by \cite{2024ApJ...977...14C}. They used the combined photometric and spectroscopic oMEGACat data to determine ages for over 8100 subgiant branch stars, using isochrones fine-tuned to the specific properties of \omc. Together with the VLT MUSE spectroscopic metallicity measurements, this work revealed a complex picture (see also the lowest row in \autoref{fig:sps_comparison}) with at least two components in age-metallicity space. One component shows a relatively tight age-metallicity relation as expected in a self-enrichment scenario. The other component is more diffuse and, on average, more metal-rich. The overall mean age in the \cite{2024ApJ...977...14C} sample is 12.06\,Gyr, with a clear difference between primordial (12.6\,Gyr) and chemically enhanced (11.6\,Gyr) stellar populations \citep{2026ApJ...998..177C}. The mean metallicity is [Fe/H]$=$-1.76. We note that \cite{2024ApJ...970..152N} found a $\sim0.07$\,dex lower mean metallicity when using red-giant-branch stars instead of subgiant-branch stars ([M/H]$=-1.612$, which corresponds to [Fe/H]$\approx$-1.83 when using the conversion described by \citealt{2023ApJ...958....8N}).

\subsection{Full Spectrum Fitting for High S/N Stacks}
\label{subsec:stellarpops}
The LVM \omc{} data allow us to test unresolved light stellar population synthesis techniques with very high S/N spectra and for different SSP template libraries.
To obtain these high S/N spectra, we co-add all spectra from non-contaminated fibers within the half-light radius  ($r_{\rm Fiber} < r_{\rm HL}$). This region contains 238 fibers, and the combined spectrum reaches a theoretical S/N of $\sim1300$; however, correlated noise likely leads to a lower true S/N value.
As there are millions of stars within the half-light radius of \omc, we expect them to fully sample the mass function, making the application of stellar population synthesis methods appropriate. To recover the stellar population properties from the stacked spectrum, we again use \texttt{pPXF}. The overall setup and the iterative masking scheme are the same as for the kinematic fits, but instead of additive polynomials, we use multiplicative polynomials of degree $N=10$. We also vary the probed wavelength range to test how this affects the populations-synthesis results.

As templates, we use the following four SSP template libraries, in the configuration that is provided in the pPXF repository\footnote{\url{https://github.com/micappe/ppxf_data}}. In all cases except XSL, we smooth the LVM spectra using Gaussian kernels to account for the lower spectral resolution of the templates:
\begin{enumerate}
    \item \texttt{galaxev/BC03} \citep{2003MNRAS.344.1000B} with Padova isochrones and Salpeter IMF. For the smoothing, we assume a uniform template resolution with FWHM of 3.0\,$\AA$ for the full wavelength range.
    \item \texttt{fsps} \citep{2009ApJ...699..486C, 2010ApJ...712..833C} with MIST isochrones and Salpeter IMF. For the smoothing, we assume a uniform template resolution with FWHM of 2.5\,$\AA$  (the resolution of the underlying MILES spectra) until $\lambda=7500\,\AA$. For longer wavelengths we use $R=\lambda/\Delta\lambda=200$.
    \item \texttt{eMILES} \citep{2016MNRAS.463.3409V} with Padova isochrones and Salpeter IMF. We assume a uniform template resolution with FWHM of 2.5\,$\AA$ over the full wavelength range.
    \item \texttt{XSL} \citep{2022A&A...661A..50V} with PARSEC/COLIBRI isochrones and Salpeter IMF. Here, no smoothing is necessary, as the templates have a higher spectral resolution ($R\sim10000$) than the LVM spectra.
\end{enumerate}
Because these libraries differ not only in their empirical stellar spectra but also in their isochrones, wavelength coverage, abundance assumptions, and spectral resolution, the comparison should be interpreted as a test of systematic sensitivity rather than as four independent measurements of a single well-defined quantity.

\begin{figure*}
    \centering
    \includegraphics[width=0.9\linewidth]{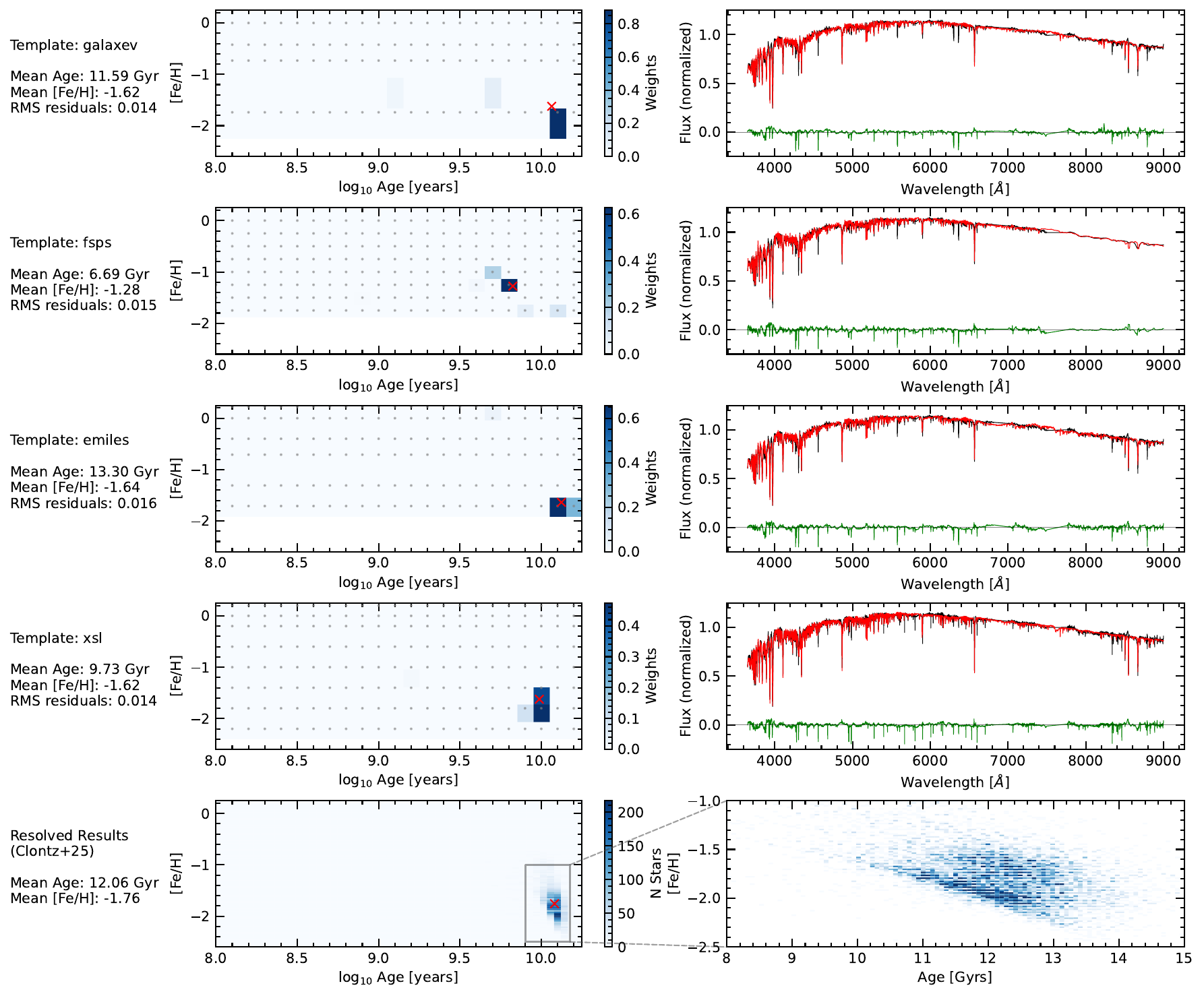}
     \caption{Comparison of different SPS library fits to the unresolved light of \omc{} with resolved results using a wavelength range of 3650\,\AA$-$9000\,\AA. The left column shows the weights in the age-metallicity grid determined by the full spectrum fit, the right column shows the best fit spectrum and the residuals. Each row corresponds to a different SSP template library. The last row shows the age-metallicity relation determined from 8000 sub-giant branch stars by \cite{2024ApJ...977...14C} with a zoom in the lower-right corner.}
      \label{fig:sps_comparison}
\end{figure*}

The best fit results and the resulting age-metallicity relations for all four template libraries for the largest probed wavelength range are shown in \autoref{fig:sps_comparison}.
The mean age and metallicity values for all configurations are shown in \autoref{tab:ssp_results}. We used the same bootstrapping approach as for the kinematics to quantify statistical uncertainties; however, given the very high S/N almost no statistical variance was detected and the reported values are clearly dominated by systematic effects.

\begin{table}[]
\centering
\caption{Mean age and metallicity values for \omc{} derived using different SSP templates and different wavelength ranges.}
\small
\label{tab:ssp_results}
\begin{tabular}{lllll}
\hline
Template & Mean Age & Mean Metal. & RMS residuals    \\
-   & Gyr      & {[}Fe/H{]}       & {[}norm. flux{]} \\ \hline
\multicolumn{4}{c}{Wavelength Range: 3650\,\AA$-$ 9000\,\AA}\\
galaxev      &  11.55  & -1.618 &  0.014 \\
fsps         &   6.90  & -1.287 &  0.014 \\
emiles       &  13.38  & -1.637 &  0.016 \\
xsl          &   9.64  & -1.617 &  0.014 \\ \hline
\multicolumn{4}{c}{Wavelength Range: 3650\,\AA$-$7500\,\AA}\\
galaxev      &  12.60  & -1.493 &  0.012 \\
fsps         &   9.00  & -1.236 &  0.014 \\
emiles       &  12.48  & -1.683 &  0.013 \\
xsl          &   9.95  & -1.624 &  0.012 \\ \hline
\multicolumn{4}{c}{Wavelength Range: 3650\,\AA$-$ 5500\,\AA}\\
galaxev      &  10.78  & -1.613 &  0.012 \\
fsps         &   7.63  & -1.251 &  0.016 \\
emiles       &  12.58  & -1.631 &  0.014 \\
xsl          &  10.71  & -1.643 &  0.013 \\ \hline
\multicolumn{4}{c}{Wavelength Range: 5500\,\AA$-$ 7500\,\AA}\\
galaxev      &  13.53  & -1.285 &  0.007 \\
fsps         &  10.16  & -1.457 &  0.008 \\
emiles       &   7.95  & -1.703 &  0.009 \\
xsl          &  11.07  & -1.577 &  0.007 \\ \hline
\multicolumn{4}{c}{Resolved results \citep{2024ApJ...977...14C}}\\
       &  12.06  & -1.76 &  --  \\ \hline
\end{tabular}
\end{table}

\subsection{Results of population synthesis}
All of the fits detect an old, metal-poor population. However, there is considerable variation in the derived mean age: When using the full wavelength range, the mean ages range from 6.69\,Gyr (fsps) to 13.30\,Gyr (eMILES) and the mean metallicities range from [Fe/H]=-1.64 (eMILES) to [Fe/H]=-1.28 (fsps). Changing the wavelength range also leads to significant variations of the mean Age (2-3\,Gyr) even with the same template library, while the metallicity seems more robust (0.1-0.2 dex).

The best agreement with the resolved observations (Mean age resolved: 12.06\,Gyr; Mean [Fe/H] resolved: -1.76) is obtained with the eMILES library and a wavelength range from 3650\,\AA$-$7500\,\AA. The overall higher [Fe/H] derived from the unresolved light spectra could be explained by the $\alpha$ enhancement in \omc. While the template libraries assume solar-scaled $\alpha$-abundances, the resolved [Fe/H] value was derived assuming [$\alpha$/Fe]$=0.3$ \citep{2010ApJ...722.1373J}.

The largest deviations among the different SSP templates are found in the comparatively young age derived with XSL and fsps and the relatively high metallicity using fsps.

These deviations are not unexpected, because old stellar populations are intrinsically difficult to age-date from integrated optical spectra: at ages above several Gyr, the spectral changes with age become subtle and can be partially degenerate with metallicity, abundance ratios, horizontal-branch morphology, and continuum-treatment choices.

While we can study and compare the mean values of the age and metallicity, we cannot resolve the spread in age and metallicity and the age-metallicity relation. This is not surprising: When looking at the age-metallicity grids of the different templates (\autoref{fig:sps_comparison}, \textit{left}) one can see that the age and metallicity resolution at the old-age end is not sufficient to resolve the variations that are visible in the resolved study. Typically, a single template spectrum contributes most of the weight ($>$0.5) to the best fit. 

\subsection{Caveats about our population synthesis tests}
Our results demonstrate that the mean properties (old, metal-poor) of \omc's stellar populations can be qualitatively recovered. However, several factors complicate both the comparison with resolved results and between the different spectral templates used. Some of these factors are general challenges of the stellar population synthesis method, others are related to the peculiar stellar populations of \omc. While we do not aim to attribute the deviations to a specific factor, or to overcome them in this work, in the following we provide a list of the main complications:
\begin{description}
    \item[Initial Mass Function (IMF): ] The IMF is a fundamental assumption in the stellar population-synthesis technique. While the four tested template libraries were all used with the same classical Salpeter IMF \citep{1955ApJ...121..161S}, this may not be an accurate representation of the present-day mass function in \omc: recent work by \cite{2023MNRAS.521.3991B} found evidence for a bottom-light mass function in massive star clusters. In addition, dynamical processes can change the radial distribution of stellar masses.
    \item[Isochrones: ] The four different template libraries used three different isochrone models, which are also different from the custom-made isochrones used in the resolved comparison dataset by \cite{2024ApJ...977...14C}. For a true one-to-one comparison, all SSP template libraries would have to be created with the same set of isochrones.
    \item[Horizontal Branch Stars: ] Another typical challenge for population synthesis of old stellar populations is the treatment of horizontal branch stars, which can bias age estimates to lower values \citep{2022MNRAS.511..341C}. \omc{} contains an extended horizontal branch, with very hot ``extreme horizontal branch" and ``blue hook" stars \citep{2018A&A...618A..15L,2023A&A...677A..86L}.
    \item[\omc's peculiar stellar populations: ] The employed stellar population templates assume a single solar-like $\alpha$ abundance and a primordial He abundance for their isochrones. This does not match the observed properties \omc, which shows alpha enhancement of [$\alpha$/Fe]$\sim$0.3 \citep{2010ApJ...722.1373J} and a significant spread of both Helium \citep[e.g.][]{2005ApJ...621..777P,2025ApJ...984..162C} and other light element abundances. In addition, the stars used to create the empirical stellar template libraries follow the [$\alpha$/Fe] abundance patterns of Milky Way field stars, adding an additional factor of systematics.
\end{description}

\subsection{Comparison with LVM-DAP Resolved Stellar Population analysis}

After reduction with the data reduction pipeline, all LVM observations are automatically processed with the Data Analysis Pipeline (LVM-DAP), which aims to fit both the emission lines and the stellar continuum for each individual fiber spectrum. A detailed description of the aims and functionality of the pipeline is given in \cite{2025AJ....169...52S}. As the LVM observes stellar populations at vastly different distances and physical scales, the number of stars sampled within a single fiber aperture can range from zero or just one up to thousands, making the universal application of classical stellar population synthesis techniques, which require a fully populated initial mass function (IMF), not always valid. Instead, the LVM-DAP follows a new approach to model resolved stellar populations (RSP), in which the spectra are decomposed into linear combinations of representative stellar spectra. Unlike classical single stellar population (SSP) templates, each RSP template has an associated probability distribution function (PDF) of physical parameters ($T_\mathrm{eff}$, $\log(g)$ ,$\mathrm{[Fe/H]}$, and $\mathrm{[\alpha/Fe]}$), allowing the pipeline to characterize the stellar content across the full dynamic range of stellar densities sampled by the LVM.

\begin{figure}
    \centering
    \includegraphics[width=1.0\linewidth]{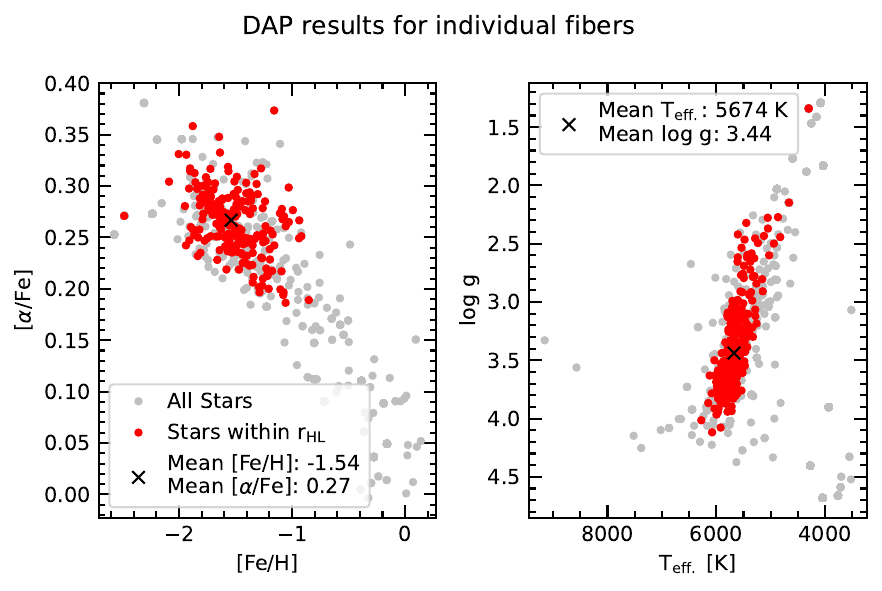}

     \caption{LVM-DAP results for the stellar populations in \omc. Grey markers show the mean value of the main quantities for all fibers, while red markers are for non-contaminated fibers within the half-light radius.}
      \label{fig:dap_results}
\end{figure}

We show the mean metallicity [Fe/H], alpha enhancement [$\alpha$/Fe], effective temperature (T$_{\rm eff.}$) and surface gravity (log\,g) as derived by the LVM-DAP in \autoref{fig:dap_results}. The results are for individual fibers of a single exposure (ID: 20836). We note that the values reported in the DAP are the mean values of a broader PDF.\\
For a fair comparison with the population synthesis results, we restrict the data to the region within the half-light radius and exclude fibers with foreground contamination (see \autoref{sec:contamination}). The results for the selected spectra show a tight distribution in [$\alpha$/Fe] over [Fe/H] space. The median metallicity is [Fe/H]$-1.54$ with an RMS of 0.23 dex, the median alpha enhancement is ([$\alpha$/Fe]$=0.27$). The metallicity value agrees within 0.2 dex with resolved studies and also with the typical mean metallicities derived via population synthesis (see \autoref{tab:ssp_results}). Also the alpha enhancement agrees well with previous high resolution studies (\citealt{2010ApJ...722.1373J} found [$\alpha$/Fe]$\sim$+0.3 for heavy alpha elements). Finally, in the surface gravity (log $g$) vs. mean effective temperature (T$_{\rm eff.}$) diagram, we can observe a vertical structure, that is matching our expectations for a spectrum dominated by red-giant-branch stars. The overall mean values are $\overline{T_{\rm eff.}}$=5674\,K  is $\overline{\rm log\,g}=3.44$. 

\subsection{Foreground extinction}
\label{subsec:extinction}
\begin{figure*}
    \centering
    \includegraphics[width=1.0\linewidth]{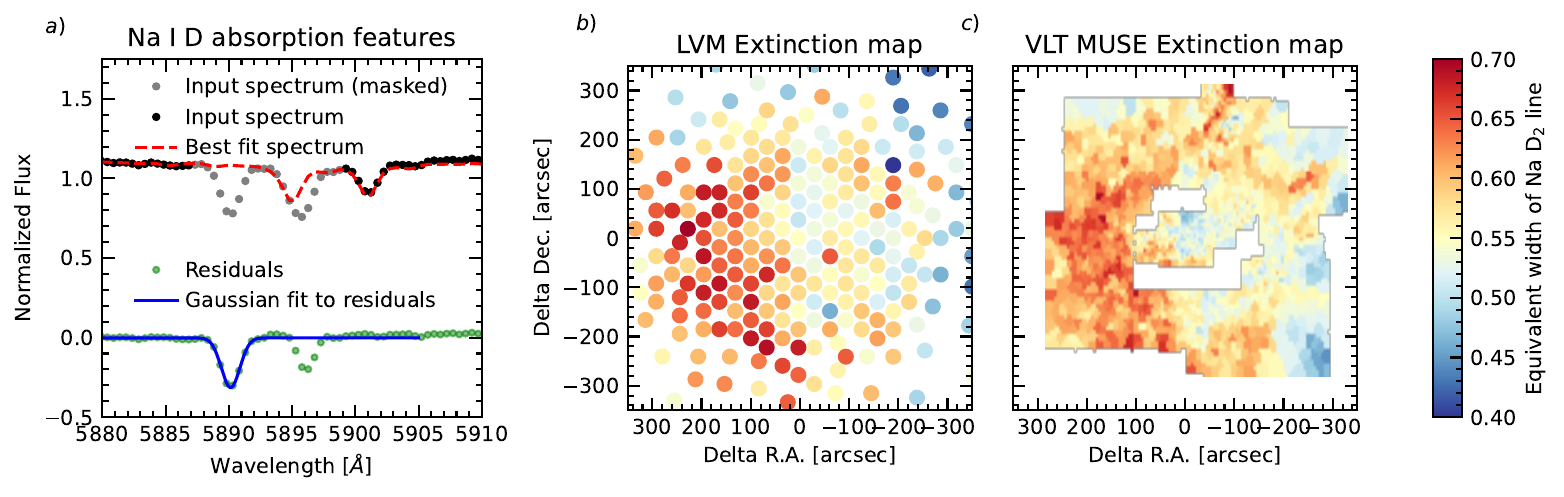}
     \caption{Study of the extinction features in the spectra of \omc. \textit{a)} shows the Na doublet region for one LVM spectrum and the best SSP fit. One can see the foreground absorption in the residuals. \textit{b)} shows the equivalent width (EW) of the foreground Na D$_2$ line for different spatial bins. \textit{c)} shows the foreground extinction measured from resolved MUSE spectra as presented by \cite{2025ApJ...994..143W}.}
      \label{fig:extinction}
\end{figure*}

Although \omc{} is located above the Galactic plane, there is still some detectable foreground extinction, which shows significant variations over the extent of the cluster. This differential extinction has been studied (and corrected) photometrically \citep{2017ApJ...842....7B,2024ApJ...970..192H,2024A&A...686A.283P} and using the resolved spectra of the oMEGACat MUSE sample \citep{2025ApJ...994..143W}. In the LVM spectra, the extinction is most notable at the Na I D lines in an intriguing way: due to the specific LOS velocity of \omc{}, the Na doublet appears as a ``triplet'' (see \autoref{fig:extinction}, \textit{a)}. The left-most absorption line is the foreground D$_2$ line, the central line is the superposition of foreground D$_1$ and \omc{} D$_2$, and the rightmost line is the \omc{} D$_1$ line. Measuring the equivalent width (EW) of the foreground Na $D_2$ line is, therefore, a direct measure of the amount of interstellar extinction. To obtain a spatially resolved extinction map, we restrict the residuals of the kinematic pPXF fit (see \autoref{sec:kinematics}) to the region ($5889\,\AA<\lambda<5893\AA$) and fit the left dip (which corresponds to the Na D$_2$ line) with a simple Gaussian. The equivalent width within $r_{\rm HL}$ ranges from 0.4 to 0.7\,$\AA$ and shows excellent agreement with similar measurements based on resolved MUSE results from \cite{2025ApJ...994..143W}, see \autoref{fig:extinction}, \textit{b) c)}. Even though exactly the same quantity is measured, the measurement setup is very different: for LVM, we are studying the Na D$_2$ EW in the residuals of SSP fits to unresolved light spectra. \cite{2025ApJ...994..143W} used spatially binned residuals of spectral fits to individual stellar spectra (using theoretical PHOENIX templates, see \citealt{2023ApJ...958....8N}). The good agreement is therefore a demonstration of the robustness of the method and of the high fidelity of the LVM spectra. At larger radii ($r>1\,r_{\rm HL}$) we note less stable behaviour of the EW measurements, likely due to the typically lower S/N of these bins and the increasing contribution of sky-residuals and foreground stars.

\section{Summary and conclusion}
\label{sec:summary}
In this work, we describe an unresolved light study of the massive star cluster \omc{} using the SDSS-V LVM wide-field integral field spectrograph.
We summarize our findings as follows:
\begin{description}
    \item[Flux calibration: ] By comparing the LVM data with the space-based Gaia catalog, we have demonstrated that the flux calibration accuracy of the standard LVM data reduction pipeline meets the requirement of 10\% for a target with a stellar continuum spectrum
    \item[Kinematic data: ]  The LVM data allow to measure the mean LOS rotation of \omc{} out to a distance of 3\,$r_{\rm HL}$. We find a rotation pattern with constant position angle, that reaches a maximum amplitude of $(8.4 \pm 0.8)$\,km\,s$^{-1}$ at $r \approx 4.7^\prime$. Comparisons with resolved MUSE measurements show good agreement, but highlight that unresolved light spectra are dominated by individual stars, which can lead to increased stochastic errors in the measurements. We make both the 1D rotation profile measured with the LVM and a combined MUSE \& LVM 2D mean velocity map publicly available to facilitate dynamical modeling.
    \item[Stellar Populations: ] Using stellar population synthesis with various commonly used SSP template libraries (galaxev, fsps, eMILES, xsl) we recover the old, metal poor stellar population of \omc{}. However, the determined mean ages and metallicities show significant variations when varying the wavelength range and the adopted template library. The detected metallicities range from [Fe/H]=-1.7 (eMILES) to [Fe/H]=-1.2 (fsps). The mean age ranges from 6.0\,Gyr (fps) to up to 13.38\,Gyr (eMILES). We discuss several potential contributors to these variations.  In contrast to the resolved study of \cite{2024ApJ...977...14C}, we are not able to resolve the spread of stellar ages and metallicities due to the step size of the age and metallicity sampling of the SSP templates.
    \item [Foreground Extinction: ] We are able to probe the amount of foreground extinction by measuring the equivalent width of the Na D$_2$ line in the LVM spectra. Our results are consistent with resolved measurements by \cite{2025ApJ...994..143W}.
    
\end{description}
The rotation map is an important ingredient for ongoing (P. Smith et al. in prep.) and future dynamical modeling efforts studying the mass distribution within \omc.

More generally, these results demonstrate the capabilities of the LVM for stellar population studies.
Potential future work includes more detailed studies of the stellar populations, e.g. by predicting unresolved light spectra for each LVM fiber making use of existing color-magnitude diagrams.
Future stellar population studies with the LVM will target the Magellanic Clouds (Zermeño et al in prep.; Ibarra-Medel et al. in prep.) and other external galaxies (Johnston et al. in prep.; Lambert et al. in prep.). Besides extragalactic targets, particularly interesting fields in our Galaxy are Baade's window (with two globular clusters and an unobscured view into the Bulge of the Milky Way fitting in one LVM pointing) and 47\,Tuc (NGC\,104). 47\,Tuc is the second brightest globular cluster, significantly more metal rich than \omc{} ([Fe/H]$\sim$-0.72, \citealt{2010arXiv1012.3224H}) and is a known fast rotator \citep{1986A&A...166..122M,2003AJ....126..772A,2017ApJ...844..167B,2018MNRAS.473.5591K,2024A&A...688A..92P}. Observations with the LVM could help to constrain its rotation in its outer regions and also would allow us to extend our stellar population tests to a higher metallicity regime.

\section*{Data availability}
The line-of-sight rotation profile (\autoref{fig:rotation_results}, \autoref{tab:rotation_profile}) and the line-of-sight 2D map (\autoref{fig:combined_map}, \autoref{tab:combined_map}) will be made publicly available at the CDS. Direct links will be added in the final published version of this manuscript.
\begin{acknowledgements}
We thank the anonymous referee for constructive feedback that lead to a clearer presentation of the results. MH thanks Francesca Pinna (IAC), Michael Hilker, Marina Rejkuba, Marco Mirabile, Magda Arnaboldi, and Felipe Lohman (ESO) for useful discussions.
AZL-A gratefully acknowledges the support provided by the Postdoctoral Program (POSDOC) of UNAM (Universidad Nacional Autónoma de México). T. Hilder is supported by an Australian Government Research Training Program (RTP) Scholarship. J.G.F-T gratefully acknowledges the support provided by ANID Fondecyt Regular No. 1260371, ANID Fondecyt Postdoc No. 3230001 (Sponsoring researcher), the Joint Committee ESO-Government of Chile under the agreement 2023 ORP 062/2023 and the support of the Doctoral Program in Artificial Intelligence, DISC-UCN. AFK acknowledges funding from the Austrian Science Fund (FWF) [grant DOI 10.55776/ESP542]. E.J.J, S.L., B.D. and S.S. acknowledge support from the ANID CATA-BASAL project FB210003. G.A.B. acknowledges the support from the ANID Basal project FB210003. E.E. and K.K gratefully acknowledge funding from the Deutsche Forschungsgemeinschaft (DFG, German Research Foundation) in the form of an Emmy Noether Research Group (grant number KR4598/2-1, PI Kreckel) and the European Research Council’s starting grant ERC StG-101077573 (“ISM-METALS"). 

Funding for the Sloan Digital Sky Survey V has been provided by the Alfred P. Sloan Foundation, the Heising-Simons Foundation, the National Science Foundation, and the Participating Institutions. SDSS acknowledges support and resources from the Center for High-Performance Computing at the University of Utah. SDSS telescopes are located at Apache Point Observatory, funded by the Astrophysical Research Consortium and operated by New Mexico State University, and at Las Campanas Observatory, operated by the Carnegie Institution for Science. The SDSS web site is \url{www.sdss.org}.

SDSS is managed by the Astrophysical Research Consortium for the Participating Institutions of the SDSS Collaboration, including Caltech, The Carnegie Institution for Science, Chilean National Time Allocation Committee (CNTAC) ratified researchers, The Flatiron Institute, the Gotham Participation Group, Harvard University, Heidelberg University, The Johns Hopkins University, L’Ecole polytechnique f\'{e}d\'{e}rale de Lausanne (EPFL), Leibniz-Institut f\"{u}r Astrophysik Potsdam (AIP), Max-Planck-Institut f\"{u}r Astronomie (MPIA Heidelberg), Max-Planck-Institut f\"{u}r Extraterrestrische Physik (MPE), Nanjing University, National Astronomical Observatories of China (NAOC), New Mexico State University, The Ohio State University, Pennsylvania State University, Smithsonian Astrophysical Observatory, Space Telescope Science Institute (STScI), the Stellar Astrophysics Participation Group, Universidad Nacional Aut\'{o}noma de M\'{e}xico, University of Arizona, University of Colorado Boulder, University of Illinois at Urbana-Champaign, University of Toronto, University of Utah, University of Virginia, Yale University, and Yunnan University.\\
This work has made use of data from the European Space Agency (ESA) mission
{\it Gaia} (\url{https://www.cosmos.esa.int/gaia}), processed by the {\it Gaia}
Data Processing and Analysis Consortium (DPAC,
\url{https://www.cosmos.esa.int/web/gaia/dpac/consortium}). Funding for the DPAC
has been provided by national institutions, in particular the institutions
participating in the {\it Gaia} Multilateral Agreement.

\end{acknowledgements}

\bibliographystyle{aa}
\bibliography{lvm_bib}

\begin{appendix}
\nolinenumbers 

\onecolumn
\section{Attenuation function at the edge of the fibers}

\begin{figure}[h]
    \centering
    \includegraphics[width=0.4\linewidth]{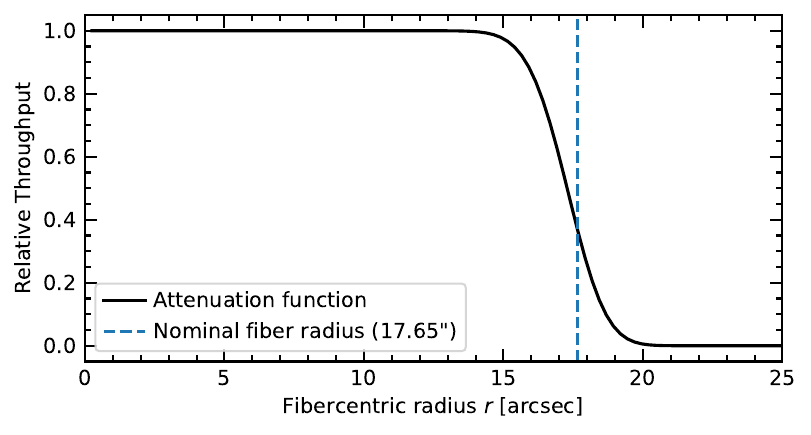}
     \caption{Attenuation function used to take into account flux losses at the edge of the LVM fibers.}
      \label{fig:attenuation}
\end{figure}

To test the absolute flux calibration of the LVM spectra, we compared the expected flux from stars from the Gaia catalog to synthetic magnitudes derived using the LVM spectra. Stars that fall on the edge of a fiber, only contribute partially to the flux of that fiber. To account for this when calculating the Gaia based fluxes, we attenuate the flux of each star with the factor $a$, depending on its distance $r$ with respect to the fiber center using the following smooth function:
\begin{equation}
    a(r) = 0.25\cdot\textrm{erfc}(\frac{r-17.67\arcsec}{\sqrt{2}\cdot1.274\arcsec})\cdot\textrm{erfc}(\frac{r-18.54\arcsec}{\sqrt{2}\cdot1.450\arcsec}),
\end{equation}
with the complementary error function $\textrm{erfc}()$ as implemented in \texttt{scipy} \citep{2020NatMe..17..261V}. A plot of the function is shown in \autoref{fig:attenuation}

\section{Published line-of-sight velocity map}
\label{appendix:published_table}

To facilitate future dynamical modeling efforts, we make a combined 2D mean LOS velocity map of \omc{} available (\autoref{fig:combined_map}). The central part of this map contains the values derived by \cite{2025ApJ...983...95H}, while the outer parts are taken from the LVM analysis presented in this work. The individual table columns are described in \autoref{tab:combined_map_columns}, exemplary table rows are presented in  \autoref{tab:combined_map}. 

\begin{table*}[h]
\centering
\caption{Description of the columns of the published LOS map.}
\label{tab:combined_map_columns}

\begin{tabular}{lll}
\hline
Column     & Unit    & Description   \\ \hline
\texttt{RA}     & degree & Mean Right Ascension of spatial bin\\
\texttt{DEC}    & degree & Mean Declination of spatial bin\\
\texttt{DX}     & arcsec  & Clustercentric coordinate x\\
\texttt{DY}     & arcsec  & Clustercentric coordinate y\\
\texttt{vlos}         & km\,s$^{-1}$ & Mean LOS velocity\\
\texttt{vlos\_corr}   & km\,s$^{-1}$ & Perspective rotation corrected mean LOS velocity\\
\texttt{e\_vlos}      & km\,s$^{-1}$ & Uncertainty of LOS velocity \\
\texttt{N} & - & Number of stars or fibers in spatial bin\\
\texttt{Inst} &  - & Instrument (MUSE in center, LVM in outskirts)\\ \hline
\end{tabular}
\end{table*}

\begin{table*}[h]
\centering
\caption{This table contains the first three and the last three rows of the published LOS velocity map data.}
\label{tab:combined_map}

\begin{tabular}{lllllllll}
\hline
RA     & DEC    & dx     & dy     & vlos         & vlos\_corr   & e\_vlos      & N & Inst \\
degree & degree & arcsec & arcsec & km\,s$^{-1}$ & km\,s$^{-1}$ & km\,s$^{-1}$ & - & -    \\ \hline
201.69605149 & -47.48172089 &    1.9 &   -7.7 &  4.12 &  4.11 &  2.44 &   70 & MUSE\\
201.69734537 & -47.47708880 &   -1.2 &    9.0 & -1.23 & -1.22 &  2.24 &   77 & MUSE\\
201.70344255 & -47.48104947 &  -16.1 &   -5.3 &  3.62 &  3.62 &  1.89 &  101 & MUSE\\
       &        &        &        &              &              &   &     & \\
201.97412766 & -47.44372939 & -674.7 &  129.1 &  6.05 &  6.44 &  0.95 &   30 &  LVM\\
201.97533720 & -47.49674831 & -677.6 &  -61.8 &  5.27 &  5.49 &  0.89 &   42 &  LVM\\
201.97749427 & -47.56730519 & -682.9 & -315.8 &  7.23 &  7.25 &  1.10 &   54 &  LVM\\ \hline

\end{tabular}
\end{table*}

\end{appendix}
\end{document}